\documentclass[sigconf,9pt]{acmart}

\usepackage[english]{babel}
\usepackage{blindtext}
\usepackage{enumitem}
\usepackage{makecell}
\usepackage[noend,ruled]{algorithm2e}
\usepackage{mathtools}
\copyrightyear{2022}
\acmYear{2022}
\setcopyright{acmlicensed}\acmConference[SIGCOMM '22]{ACM SIGCOMM 2022 Conference}{August 22--26, 2022}{Amsterdam, Netherlands}
\acmBooktitle{ACM SIGCOMM 2022 Conference (SIGCOMM '22), August 22--26, 2022, Amsterdam, Netherlands}
\acmPrice{15.00}
\acmDOI{10.1145/3544216.3544258}
\acmISBN{978-1-4503-9420-8/22/08}

\begin{document}

\title{Underwater Messaging   Using  Mobile Devices}

%\author{Paper \#472, 12 pages + references}

\author{Tuochao Chen,$^\diamond$ Justin Chan$^\diamond$ and Shyamnath Gollakota}
%\email{{ctc1998,jucha, gshyam}@cs.washington.edu}
\affiliation{%
  \institution{$^\diamond$Co-primary student authors}
    \institution{Paul G. Allen School of Computer Science \& Engineering}
  \institution{University of Washington, Seattle, WA, USA}
}

 \email{underwatermessaging@cs.washington.edu}

\renewcommand{\shortauthors}{Chen, Chan and Gollakota}

\newcommand{\xref}[1]{\S\ref{#1}}

\newcommand{\squishlist}{\begin{itemize}[itemsep=1pt,parsep=2pt,topsep=3pt,partopsep=0pt,leftmargin=0em, itemindent=1em,labelwidth=1em,labelsep=0.5em]}
\newcommand{\squishend}{\end{itemize}}

\newcommand{\squishenum}{\begin{enumerate}[itemsep=1pt,parsep=2pt,topsep=3pt,partopsep=0pt,leftmargin=0em,listparindent=1.5em,labelwidth=1em,labelsep=0.5em]}
\newcommand{\squishsubenum}{\begin{enumerate}[itemsep=1pt,parsep=2pt,topsep=0pt,partopsep=0pt,leftmargin=0em,listparindent=1.5em,labelwidth=1em,labelsep=0.5em]}
\newcommand{\squishenumend}{\end{enumerate}}

%\begin{abstract}

\begin{abstract}  Since its inception, underwater digital acoustic communication has required custom  hardware that neither has the economies of scale nor is pervasive.  We present the first acoustic system  that brings underwater messaging capabilities to existing  mobile devices like smartphones and smart watches. Our software-only solution leverages audio sensors, i.e., microphones and speakers,  ubiquitous in today's devices to enable acoustic underwater communication between mobile devices.  To achieve this, we design a communication system that in real-time adapts to differences in frequency responses across mobile devices, changes in multipath and noise levels at different locations  and  dynamic channel changes due to mobility. We evaluate our system in six different real-world underwater environments with depths of 2-15~m in the presence of boats, ships and people fishing and kayaking. Our results  show  that our system can in real-time adapt its frequency band and achieve  bit rates  of 100~bps to 1.8~kbps and a range of 30~m. By using a lower bit rate of 10-20~bps, we can further increase the range to   100~m.  As smartphones and watches are increasingly being used in  underwater scenarios,  our software-based  approach  has the potential to make  underwater messaging   capabilities widely available  to anyone with a mobile device.

\begin{center}Project page with open-source code and data can be found here: 
{\textcolor{blue}{{{\url{https://underwatermessaging.cs.washington.edu/}}}}}\end{center}
    
\end{abstract}




\maketitle

\section{Introduction}

%Each year, there are millions of scuba diving participants in the  United States alone~\cite{stats}  and tens of millions more take part in underwater activities like  snorkeling in lakes, rivers and oceans~\cite{snorkeling}.

Each year, tens of millions take part in underwater activities like  snorkeling in lakes, rivers, and oceans~\cite{snorkeling}. Millions more participate in recreational scuba diving in the  United States alone~\cite{stats}. Underwater  exploration is also important  in observation of marine life, water pollution   and to document the biological, geological, and archaeological aspects of underwater environments~\cite{exploration}. Effective  communication  during  these  underwater   activities is critical for safety and   navigation~\cite{signal1,signal2,signal3}. Hand signals are commonly used to  communicate intention,  convey  direction and maintain safety. In addition to the  10-20 signals commonly used in  recreational settings~\cite{signal1}, professional divers use  more than 200 hand signals to communicate with each other or with surface members of the dive team~\cite{signal2}. This includes information about  oxygen level, aquatic life or operations that require cooperation~\cite{signal3}. Given the number of hand signals and their visual nature, however, this mode of  communication is  ineffective in low-visibility scenarios (e.g., turbid  waters) and is  limited in its  communication range and reliability. 

To address this problem, commercial   efforts have  designed   hardware  that enables   two-way text messaging to send from a set of  pre-defined messages or SOS beacons~\cite{commercial}. DARPA also initiated the AMEBA effort  to build custom  hardware that would enable  divers to communicate at a low bit-rate   with each other via text messages or with nearby  relay buoys~\cite{darpa}.  These prior efforts  require custom hardware that is neither ubiquitous nor  has the economies of scale.

\begin{figure}[t!]
    \includegraphics[width=.38\textwidth]{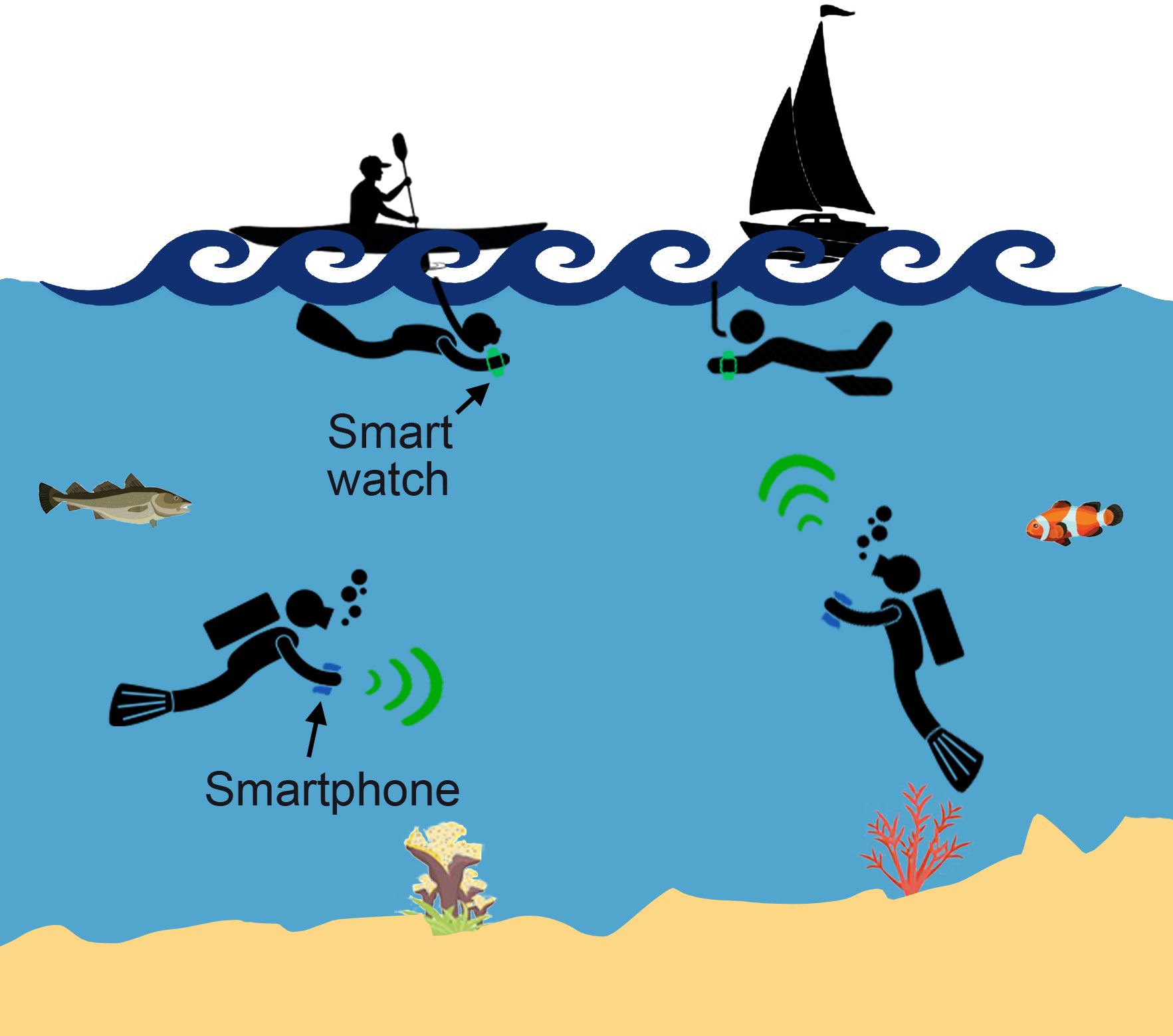}
    \vskip -0.15in
    \caption{Mobile devices for underwater messaging.}
\vskip -0.22in
\label{fig:fig1}
\end{figure}
In this paper, we take a different approach and explore if one can  enable underwater messaging  capabilities on mobile devices like smartphones and smartwatches.
Smartphones are increasingly  being used with diving-proof  cases (\$30-40)  for underwater  photography, video logging and in lieu of a dive computer~\cite{case1,case2}. {These diving-proof smartphone cases are rated to work at depths of 15-40~m depending on their price~\cite{phone1,phone2}.}  The latest   smartwatches are also  water-resistant and can be  used  during  shallow water activities like  snorkeling~\cite{watch1,watch2}.
Ideally, anyone with a mobile device should be able to  download a software app  and communicate   underwater during  snorkeling, recreational diving or scientific exploration, without  additional hardware. Such an approach would  leverage the ubiquity of mobile devices to  democratize underwater communication   and make it available to anyone with a mobile device.\footnote{Recent reports note  Apple's R\&D interest in enabling iPhones to communicate in underwater environments~\cite{apple1, apple2}.}

We present {\it AquaApp}, the first software-only solution that enables underwater acoustic communication and networking   on commodity mobile devices. While the most common communication modality on mobile devices is to use radios (e.g., Wi-Fi and LTE), these frequencies are not suitable for underwater communication --- 2.4~GHz Wi-Fi signals can attenuate as much as 169~dB per meter in seawater~\cite{wifi,shrimp}. Our experiments show that two smartphones separated by just a few inches in fresh water  could not connect using Wi-Fi or Bluetooth. Our solution instead is to  leverages  audio sensors, i.e., microphones and speakers, that are ubiquitous in today's devices to enable acoustic  underwater communication. In contrast to RF signals, acoustic signals have much   better propagation properties underwater~\cite{wifi}. %Further, recent trends show an increase in the number of  microphones in smartphones. This provides  opportunities for using receiver diversity to improve the  reliability of underwater  communication. 

\begin{figure}[t!]
    \includegraphics[width=.4\textwidth]{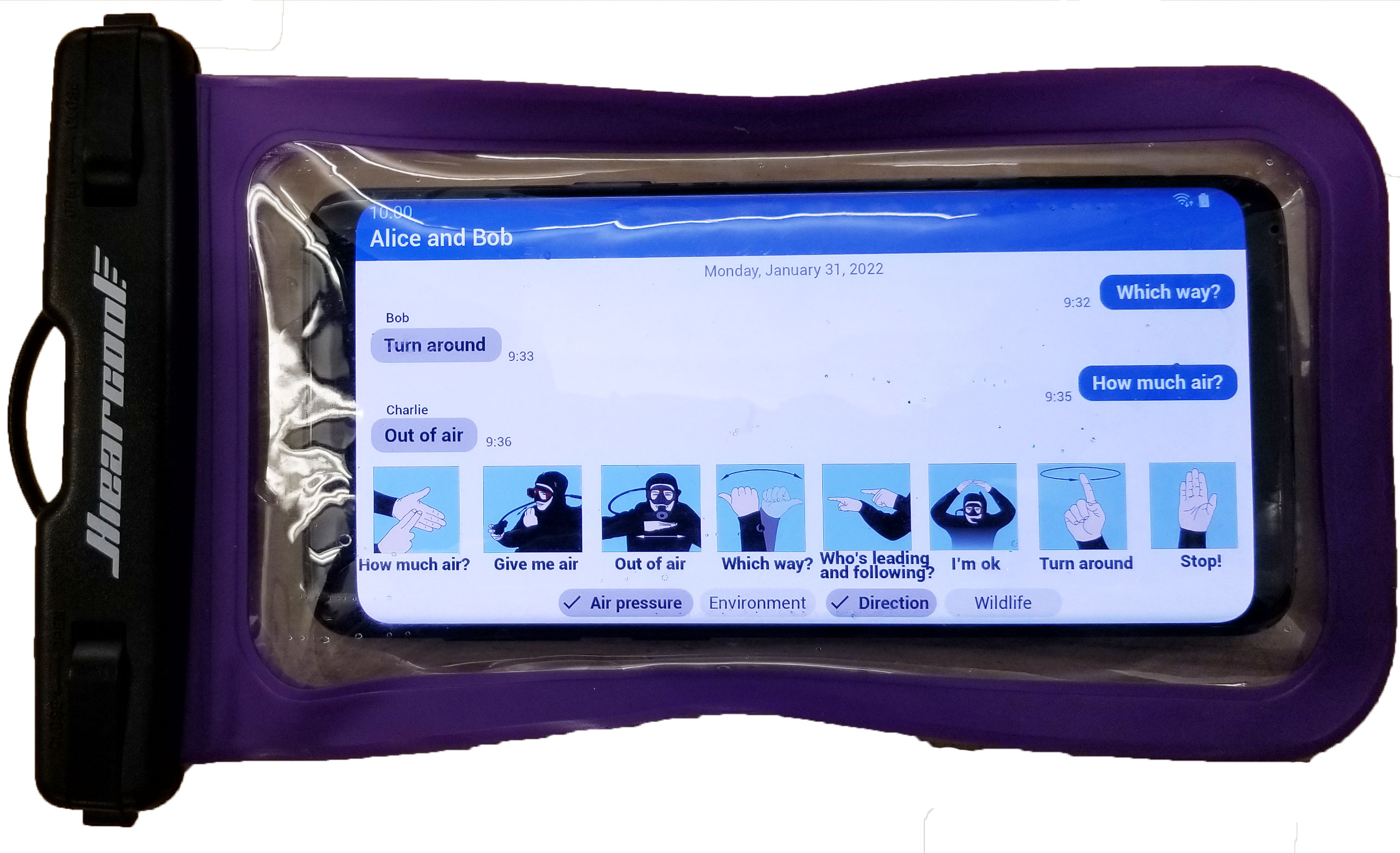}
    \vskip -0.15in
\caption{{\bf User interface of our messaging system  on a phone  {within a water-proof case}.}}
\vskip -0.15in
\label{fig:ui}
\end{figure}

Repurposing microphones and speakers on commodity mobile devices for underwater communication is challenging for three   reasons: (1) Unlike custom  hydrophone  hardware that is specially designed to operate underwater, acoustic sensors in smartphones and watches are designed for  in-air operation  and the hardware components used can vary  across device  manufacturers. This results in   different frequency responses across devices (Fig.~\ref{fig:devices}a,b). The transmit power on these mobile devices is also typically  limited compared to underwater hydrophones. In addition, the severe multipath in underwater scenarios  can result in the signal strength varying  by as much as 10-20~dB  within a few kHz. (2) Even with   the same smartphone model used  across users, the SNR   profiles on the forward and backward paths can be different (Fig.~\ref{fig:devices}c,d), resulting in the need to use  different frequency bands on the two paths. 
(3) The  bit rate  can vary an order-of-magnitude from 100~bps to 1.8~kbps  with distance and multipath, requiring an adaptation algorithm to minimize packet error rate. However, mobility inherent to   diving and snorkeling results in  varying channel across packets, making  adaptation challenging. This  requires a  real-time protocol to pick the correct  bit rate before transmitting data, without incurring significant overhead.

\begin{figure*}[t!]
    \includegraphics[width=\textwidth]{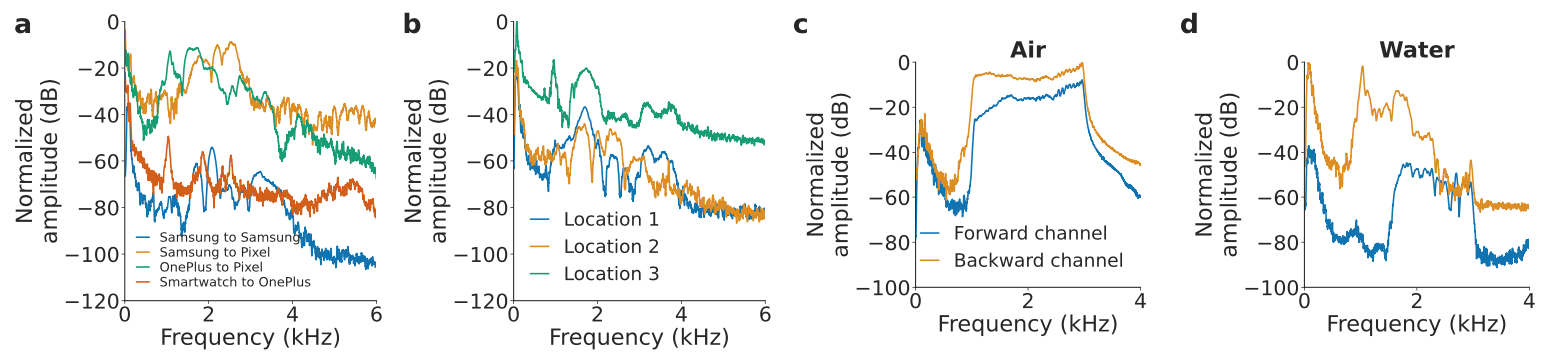}
    \vskip -0.1in
\caption{(a,b) Frequency selectivity underwater in a lake at a 5~m distance  across devices and locations. (c,d) Channel reciprocity on the forward and backward path in (c) air and (d) water separated by 2~m.}
 \vskip -0.15in
\label{fig:devices}
\end{figure*}

At a high level, we use orthogonal frequency division multiplexing (OFDM) to communicate underwater between mobile devices. Our real-time system adapts the acoustic  frequencies  used to encode data in each packet transmission, as a function of frequency response,  distance and signal-to-noise ratio (SNR) of each frequency bin. This ensures that when the devices are  a few meters apart and have a high SNR, the system adapts in real-time to use all the OFDM bins to encode data and achieve a high bit rate. On the other hand, it uses a smaller frequency band as the distance increases. This ensures that more power is allocated to a right set of frequencies, thus, increasing the SNR  and the packet delivery rate.

Our design has three key components:
\squishlist
\item {\it Post-preamble   feedback.} In  high-data rate systems (e.g., Wi-Fi), rate adaptation is performed across packets since  packet sizes are less than a millisecond, and the channel coherent time is more than a few packets. In contrast, our underwater acoustic system is low rate making packets  much longer and resulting in  the channel changing   between consecutive packets. To do this, we perform per-packet adaptation by splitting the preamble/header and data portions  of each packet. Say Alice wants to send a packet to Bob. In our design, Alice first broadcasts the preamble and header with Bob's address and stays silent for Bob's feedback before transmitting the data portion of the packet  (Fig.~\ref{fig:protocol}).  Bob  estimates the SNR for each OFDM bin using the preamble, runs a  frequency adaptation algorithm and in real-time sends information back to Alice,  embedded in a single OFDM symbol.  Alice   uses this feedback  to transmit the data portion of the packet  using  the right frequency band. %provided by Bob. 
\item {\it Frequency band   adaptation.}
 Per-frequency rate adaptation is ideally performed using the    water-filling algorithm to allocate different power and modulation to each OFDM  bin~\cite{fara}. In a low-data rate system, however,  conveying  fine-grained feedback  about  60 OFDM bins requires at least O(60) bits which is a significant overhead. To minimize the feedback Bob sends back to Alice,  we design a low-overhead  frequency band adaptation algorithm (\xref{sec:adapt}). At a high level, we first compute   the SNR in each bin  using the received preamble. If not all the OFDM  bins are above an SNR threshold, we drop  the lowest SNR bin and  reallocate power to the remaining bins. We  repeat this process until we find the largest contiguous band where all frequencies are above the SNR threshold. Bob  sends back  only information about the start and end frequencies, $f_{start}$ and $f_{end}$, of this contiguous band. 
 
 \item {\it Feedback encoding method.} 
 Bob encodes this feedback   using a single OFDM symbol.
 Our encoding method   allocates  all the power to  the  two OFDM bins corresponding to the start and end frequencies, $f_{start}$ and $f_{end}$, output by our frequency band  adaptation algorithm. Alice can   extract this information  by performing a sliding window and  picking the top-2 OFDM bins with the highest power. This is a reliable encoding method   since all the transmit power is being allocated to these two OFDM  bins. Alice then uses this   band to send data  by setting the OFDM bins outside it to zero.

\squishend

We implemented our software system in real-time on the Android platform so it can be used with  various smartphone models and  smart watches.  Since underwater multipath  can have a large delay spread, we implement time-domain equalization to reduce the cyclic prefix duration to only 7\% of the OFDM symbol. To address mobility, we  use differential coding across consecutive OFDM symbols to reduce errors (\xref{sec:data}). We also design a carrier sense based MAC protocol to support a network of multiple underwater mobile devices (\xref{sec:mac}). Using our communication system, we implement a messaging app where  users can  transmit  one of the 240 common messages. Finally, by using a lower bit rate, we design a longer-range SoS beacon messaging system.

We evaluate our system in six different real-world underwater environments. Our findings are as follows.
\squishlist
\item Our system adapts the frequency band to achieve bit rates of 100~bps-1.8~kbps at distances up to 30~m. The bit rate scales with distance and multipath. By reducing the bit rate to 10-20~bps, our system can   increase its  range to  100~m.
\item { Our real-time adaptation algorithm reduces the average packet error rate (PER) from 38-70\% to 3\% compared to fixed bandwidth schemes  across 5 to 30~m.}
\item {Our system achieves a  PER of 4\% and 7\%  in the presence of slow and fast motion. Further,   it adapts its frequency band to operate reliably with different phone orientations.}
\squishend

\vskip 0.05in\noindent{\bf Contributions.}  The last few decades have shown that software can  bring technology to the masses more rapidly than custom hardware. {We present the first acoustic-based  system that enables underwater messaging   on commodity mobile devices, using only software.} To this end, we designed a communication system that in real-time adapts to variations in frequency responses across mobile devices, changes in multipath due to mobility and SNR variations from severe frequency diversity, to minimize packet error rate.  We  evaluated our system in underwater environments  and built a messaging app that allows users to send messages and   SoS beacons. Finally, by making our system code publicly  available at publication,  we believe that this work has the potential to make underwater communication technology  accessible to everyone with a mobile device, by just  downloading software.

\begin{figure}[t!]
\vskip -0.05in
    \includegraphics[width=.42\textwidth]{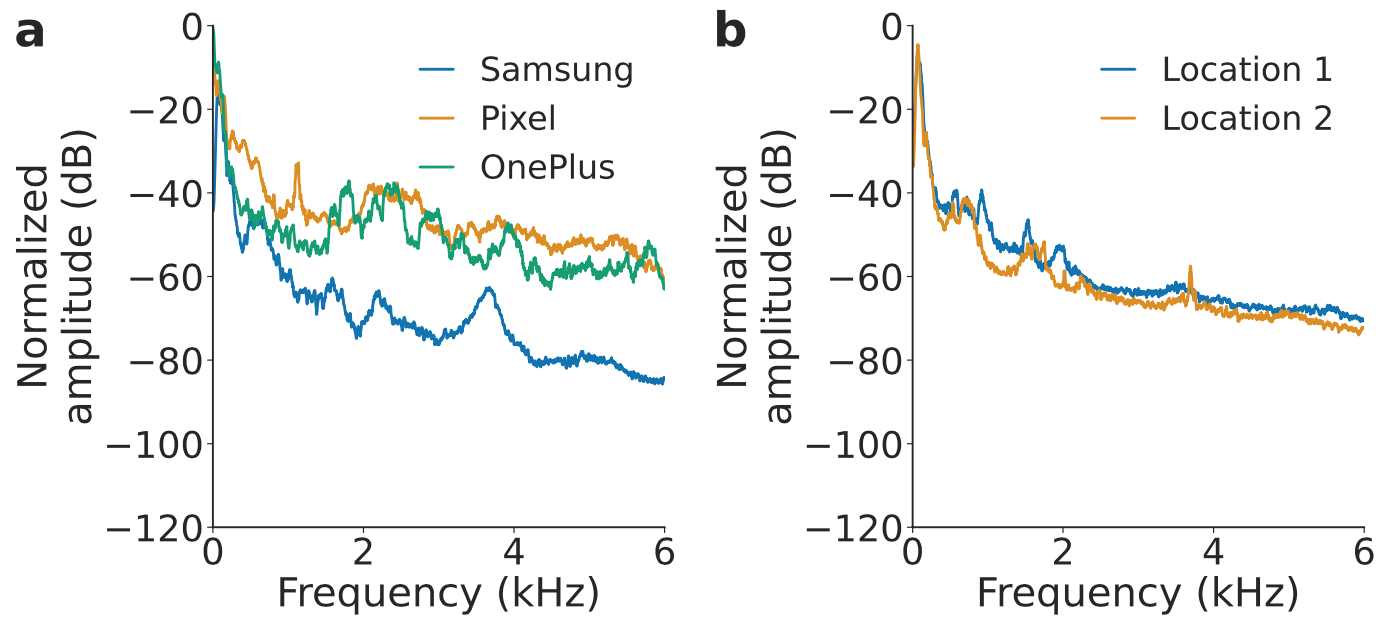}
     \vskip -0.15in
\caption{Underwater ambient noise measurements. Normalization is done  across plots in the same graph.}
 \vskip -0.15in
\label{fig:noise}
\end{figure}
\section{System Design}
We first characterize the properties of acoustic signals from mobile devices in water. We then present our real-time protocol as well as our data encoding and decoding algorithms. We open-source our code and data at {\textcolor{blue}{{{\url{https://underwatermessaging.cs.washington.edu/}}}}}.

\subsection{Characterizing mobile devices in water}  \label{sec:char}
We characterize the acoustic properties of commodity smart devices (e.g., phones, watches) in underwater scenarios.

\vskip 0.05in\noindent{\it Frequency selectivity.} As the speakers and microphones differ between smart devices, the frequency selectivity of an acoustic signal can vary between different transmitting and receiving phones. To evaluate this, we selected four different smart devices: Samsung Galaxy S9, Google Pixel 4, OnePlus 8 Pro, and Samsung Galaxy Watch 4.  We placed each device pair underwater  5~m away from each other. We placed each smartphone in a waterproof pouch~\cite{pouch} and submerged the device to a depth of 1~m in a 2~m deep lake. The transmitter sends a 1--5~kHz chirp with a duration of 500~ms. Fig.~\ref{fig:devices}a shows  that the frequency response varies between device pairs. It is uneven and exhibits deep notches with  the frequencies where notches occur varying across device. The plot also shows that  the frequency response  diminished above 4~kHz, which suggests that acoustic communication above this frequency on mobile devices may be challenging.

Next, we fix the transmitting and receiving smart devices to both be a Samsung Galaxy S9. We repeat the same experiment as before at a distance of 10~m. Fig.~\ref{fig:devices}b shows the variation in frequency response as a result of the multipath characteristics at the different locations. Specifically, we observe that multipath causes the notches of the response to occur  at different frequencies. Thus the frequencies  ideal for underwater  communication may vary with location.

\vskip 0.05in\noindent{\it Channel reciprocity.} Next, we analyze the frequency response of the forward and backward channel in air and underwater. We use two smartphones of the same model (Samsung Galaxy S9) and measure the frequency response of a 1--3~kHz chirp with a 1~s duration. The first phone is set to send a chirp to the second phone, and two seconds later, the second phone sends a chirp back to the first phone. We separate the phones by 2~m and perform the measurement over the air and underwater. In Fig.~\ref{fig:devices}c we see that the frequency response of the chirp sent over the air is similar across both phones.  However, the frequency response of the chirp underwater in Fig.~\ref{fig:devices}d differs significantly. This suggests that the optimal frequencies for communication are different for the forward and backward channel. So an explicit feedback signal would need to be sent from the receiving device to the transmitting device.

\vskip 0.05in\noindent{\it Ambient noise.} 
Finally, we measure underwater  ambient noise   as recorded on different smart devices  for five seconds at the same location. Fig.~\ref{fig:noise}a shows the ambient noise across these devices for different frequencies. Each noise profile is normalized to the maximum amplitude across all measured frequencies. The plot shows that across devices, the amplitude of noise is high below 1~kHz. Furthermore, noise can also be seen at higher frequencies up to 4.5~kHz. In our measurement location, underwater noise was due to the sound of water flowing and the movement of air bubbles. Additionally, the plot shows that the noise profiles vary across smartphones.  We also measured the level of ambient noise at different locations using a Samsung Galaxy S9.  Fig.~\ref{fig:noise}b shows that the level of noise between 0--6~kHz can vary by 9~dB between locations. This is expected as the amount of water flowing or other acoustic interference will vary across locations. Across both these experiments, the high noise level below 1~kHz suggests that acoustic communication in that frequency range may be challenging. This suggests that we need to measure both signal strength and noise across frequencies to determine the right frequencies. % for use in underwater  communication with mobile devices.

\vskip 0.05in\noindent{\bf Design requirements.}  So our requirements are:

\squishlist
\item \textit{Works across smart devices with different frequency responses.} Our system  should work across speakers and microphones on smart devices with  different frequency responses. 
\item \textit{Robust to multipath across  different locations and distances.} Underwater environments can have challenging multipath due to reflections from the surface, floor and  from the coast. Our system should be robust to these effects.
\item \textit{Tolerates mobility.} Strong water currents or waves can cause devices to  drift away quickly or toward each other within a few seconds, resulting in Doppler shifts. Our system should be able to work in these real world environments.
\item \textit{Adapts to different noise profiles across  environments.} The underwater environments  can suffer from significant noise from   ships, boats and animals. Loud  sounds outside the water such as airplanes or helicopters  also contribute to the noise.
\squishend

\subsection{Post-preamble feedback protocol}\label{sec:protocol}

%First, in Section2, we find that the forward (from Alice to Bob) and backward acoustic channel (from Bob to Alice) underwater is significantly different. This implies that we are unable to apply the channel reciprocity for reverse channel estimation. Hence, if Alice want to know the forward channel characteristics from Alice to Bob, it requires the feedback signal from Bob about the forward channel information.

%Secondly, the underwater acoustic channel is dynamically changing, due to the motion of device and dynamic environment itself (e.g. the turbulence and wave). Unlike other high-rate traditional OFDM system (e.g. WiFi), our acoustic-based underwater communication system has lower data rate and longer packet duration. This means that even in two consecutive packets, the frequency selectivity of channel may change.  Hence, we should apply the adaptation for each packet to be transmitted. Furthermore, we need to try our best to reduce the overhead of feedback-based channel estimation.

Fig.~\ref{fig:protocol} shows  our protocol that splits the packet into preamble/header and data.  Alice  first sends the preamble  and the receiver ID. Alice remains silent for feedback from Bob but keeps its OFDM symbol timer ON. The preamble is composed of eight  OFDM symbols from  1 to 4 kHz. When Bob detected this preamble, we first check the ID. It then runs our SNR estimation algorithm on the preamble for each subcarrier between 1-4 kHz, which it uses in our frequency band adaptation algorithm. Instead of a fine-grained adaptation for each  bin, we select a contiguous band and send back only  the start and end frequencies of this band to reduce  overhead. Alice uses this information to encode bits  within this selected frequency band and  transmits  data  to Bob. To do this, Alice transmits the  data symbols at the beginning of the next OFDM time interval as determined by its OFDM symbol clock.  This ensures that the preamble  synchronization performed by Bob can also be used for data. The first OFDM symbol in the data transmitted by Alice however is a known training symbol to track changes in channel since preamble transmission. In practice, the post-preamble silence period  at Alice is assigned zero values to keep the  speaker buffer full. This  ensures that the symbol timing can be maintained for data symbols.  {Bob uses   cross-correlation and energy detection  in every OFDM symbol interval to detect the arrival of the first known  data symbol from Alice. Bob can use the preamble to synchronize the data symbols since the  propagation time for preamble and data symbols is similar. Note that safe human motion under water is usually lower than $1-2m/s$~\cite{wojtkow2017biomechanics} and the time interval between the preamble and data transmission in Alice side is several OFDM symbols (including feedback propagation, processing time). Assuming the interval is 5-symbol,  the change in the propagation time  is only 0.6\% of the OFDM symbol duration.}

\begin{figure}[t!]
    \includegraphics[width=.44\textwidth]{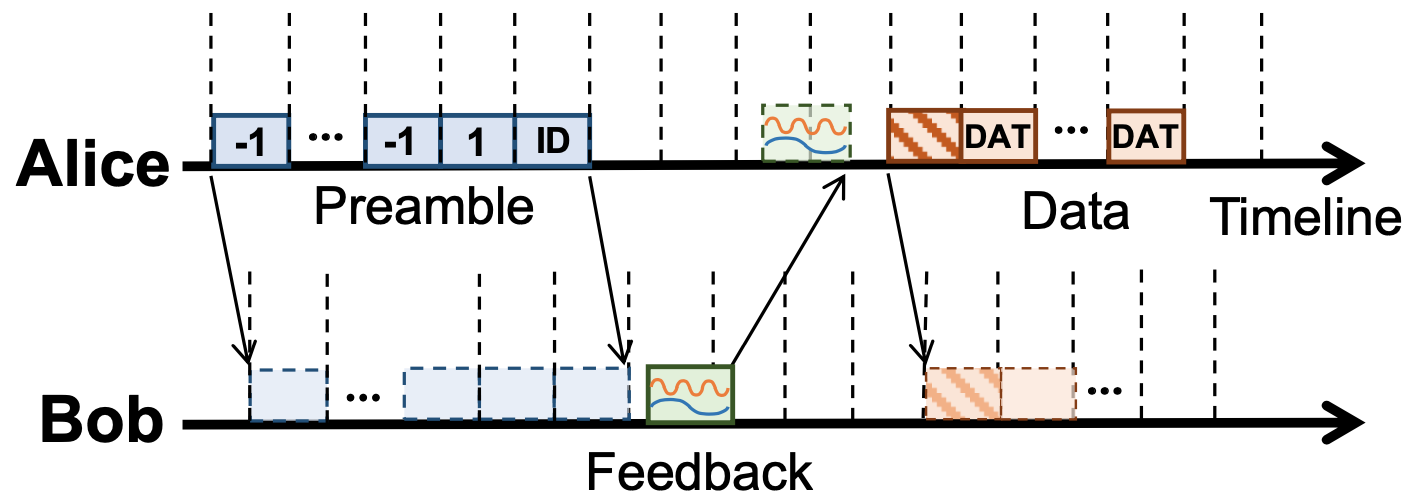}
    \vskip -0.15in
\caption{{\bf Protocol sequence diagram.}}
\vskip -0.15in
\label{fig:protocol}
\end{figure}

\subsubsection{Preamble design}
 The preamble has three  purposes: packet detection, symbol synchronization, and channel estimation.  For a real-time system, the preamble design and detection algorithm have two main requirements: detection robustness and low computational burden.   Linear frequency modulation (LFM) signals and cross-correlation-based detection are  proposed for   underwater  communication~\cite{xie2009implementation, 8190891, thottappilly2011ofdm}. However, in our  real-world experiments,  LFM was not robust enough at long distances or with severe multipath. We instead use a data-aided approach~\cite{nasir2010performance}  where we fill the OFDM bins with a CAZAC sequence. These sequences have a very good auto-correlation property and have unit Peak to Average Power Ratio (PAPR)~\cite{wen2006cazac}.  Further, the good autocorrelation properties of a CAZAC sequence also makes it suitable to be used for channel estimation~\cite{zhang2020endophasia}.
We  concatenate  eight such identical OFDM symbols  and multiply each with a  PN sequence with different signs ([-1, 1, 1, 1, 1, 1, -1, 1]), to provide steeper fall off to the correlation timing metric and alleviate the side-lobes problem for a more robust detection and accurate synchronization~\cite{wen2006cazac}. 

\textit{Preamble detection and synchronization.}  Our detection algorithm is composed of two parts: coarse detection and fine-grained detection.
The coarse detection algorithm applies  cross-correlation between the received signal and  preamble. In the presence of a preamble, this results in a correlation peak. However,  the cross-correlation peak varies with  SNR and spiky noise like underwater bubbles could also cause a very high correlation peak. To address this, we also use sliding correlation~\cite{nasir2010performance} where instead of calculating the correlation between the transmitted preamble and received signal, we apply a sliding window  on the received signal and divide the sliding windows into 8 segments each the length of an  OFDM symbol. Then we multiply each segment by the PN sequence and calculate the correlation between the  two nearby segments. Finally, we sum them up and divide the sum by the received energy  within this sliding window. When the real preamble arrives, the sliding correlation would have a high peak ($> 0.6$). The  peak height of this normalized  sliding correlation is not sensitive to SNR changes. Moreover,  spiky  noise is unlikely to have this  specific  encoded data pattern and thus its sliding correlation value is  low ($<0.2$).

\begin{algorithm}[t!]\label{algo:alg3}
\caption{Frequency band  selection algorithm} 
\DontPrintSemicolon
\For{$L \leftarrow N_0$ to $1$} {
\For{$m \leftarrow 0$ to $N_0-L$} {
$\displaystyle SNR_{min}= \min_{m \leq k < m +L} \{SNR_k+ \lambda\cdot 10 log_{10}(\frac{N_0}{L})\}$\\
\If{$SNR_{min} > \epsilon_{SNR}$} {
$n \gets m+L-1$\\
\Return $(m, n)$
}
}
}
% \vspace{-0.1sin}
\end{algorithm}
% \vskip -0.15in

% \begin{algorithm}[t!] 
% 	\caption{Frequency band  selection algorithm} 
% 	\label{alg3} 
% 	\begin{algorithmic}
%         \FOR{$L \leftarrow N_0$ to $1$} 
% 		\FOR{$m \leftarrow 0$ to $N_0-L$} 
% 		\STATE $\displaystyle SNR_{min}= \min_{m \leq i < m +L} \{SNR(i)+ \lambda\cdot 10 log_{10}(\frac{N_0}{L})\}$
% 		\IF{$SNR_{min} > \epsilon_{SNR}$}
% 		\STATE $n \gets m+L-1$
% 		\RETURN (m, n)
% 		\ENDIF
%         \ENDFOR 
% 		\ENDFOR 
% 	\end{algorithmic} 
% \end{algorithm}
However, the  sliding correlation method increases  computational burden. So we perform  the sliding correlation on the candidate signals  only after the coarse detection. In addition, we increase the step size of sliding correlation to 8 to balance the computational burden and synchronization resolution. When a valid preamble is detected, we select the peak index in the sliding correlation curve as the beginning of the preamble for synchronization.  Even if our preamble structure can improve the peak prominence in sliding correlation, a synchronization offset is still inevitable, which may affect OFDM data transmission quality~\cite{nasraoui2015advanced}. As described later,  we use time-domain equalization in our decoder and a cyclic prefix in our encoder to address this issue.
%$$M(n) = \frac{\sum_{k=0}^6 p(k)p(k+1) \sum_{m=0}^{Ns} R^*(n+k*Ns+m)R(n+(k+1)*Ns+m)}{\sum_{k=0}^7 \sum_{m=0}^{Ns} R^*(n+k*Ns+m)R(n+k*Ns+m)} $$
%Specifically, for each preamble candidate $R(n)$ from the coarse detection, we first apply the FIR bandpass filter ([1KHz, 4KHz]) to filter out the noise level out of preamble bandwidth. Then we calculate the sliding correlation using the below equation

%%\begin{figure}[t!]
 %%   \includegraphics[width=.47\textwidth]{./figs/packet.png}
%%\caption{{\bf The packet structure for preamble, feedback signal and data signal.}}
%%\label{fig:protocol}
%%\end{figure}

\subsubsection{Frequency band adaptation algorithm}\label{sec:adapt}
Our frequency band  adaptation algorithm is composed of two main steps. 

\textbf{SNR estimation per frequency bin.} 
We use the 8 OFDM symbols in the preamble  to estimate the channel. We compute the SNR for each subcarrier by applying  frequency-domain channel estimation. Specifically, we denote the transmitted data in subcarrier $k$ of the eight preamble symbols by a vector $x(k)$ and the received data in subcarrier $k$ of 8 training symbols by a vector $y(k)$. We apply a minimum mean square error (MMSE) estimator to compute  the channel response $H(k)$ for each subcarrier $k$. We then compute the SNR in the $k^{th}$ bin as, $SNR_k = 20log_{10}\frac{\|H(k)x(k)\|_2}{\|y(k)-H(k)x(k)\|_2}$
%$SNR_k = 20log_{10}\frac{\|H(k)x(k)\|}{|y(k)-H(k)x(k)|}$.

\textbf{Frequency band selection.} 
Our goal  is to find the optimal frequency range based on the SNR distribution between  1-4~kHz. The basic idea  is that we drop the bin with the  lowest SNR and reallocate power to the remaining bins until the SNR in all remaining bins surpass the preset SNR threshold. Say,  there are totally $N_0$ bins between 1-4~kHz, and the estimated SNR in the $k^{th}$ bin is $SNR_k$. When we only select  bins between $m$ and $n$, the power in the discarded bins could be re-allocated to the remaining bins. Hence, the SNR value in the remaining bins  increases by $10  log_{10}(\frac{N_0}{L})$, where $L=n-m+1$. 
So, our optimization problem is as follows,
\begin{align*}
&\max_{m,n}\quad L = n-m +1\\
& \begin{array}{r@{\quad}r@{}l@{\quad}l}
s.t.&  &SNR_k + \lambda \cdot 10 log_{10}(\frac{N_0}{L})> \epsilon_{SNR}, \forall k\in[m,n]\\
\end{array}
\end{align*}
where $\epsilon_{SNR}$ is the preset SNR threshold (in our  implementation we set $\epsilon_{SNR}$ to 7). $\lambda$ is a conservative factor between 0 to 1 (we select it empirically to 0.8), since in real-world implementation the power re-allocation is not exact. We set, $\epsilon_{SNR}$ and $\lambda$, a bit conservatively since  we need to  account for  imperfect SNR estimation  and because the  channel may still change due to mobility.
 $m$ and $n$ from the above optimization  give us  $f_{begin}$ and $f_{end}$.

\begin{figure}[t!]
    \includegraphics[width=.38\textwidth]{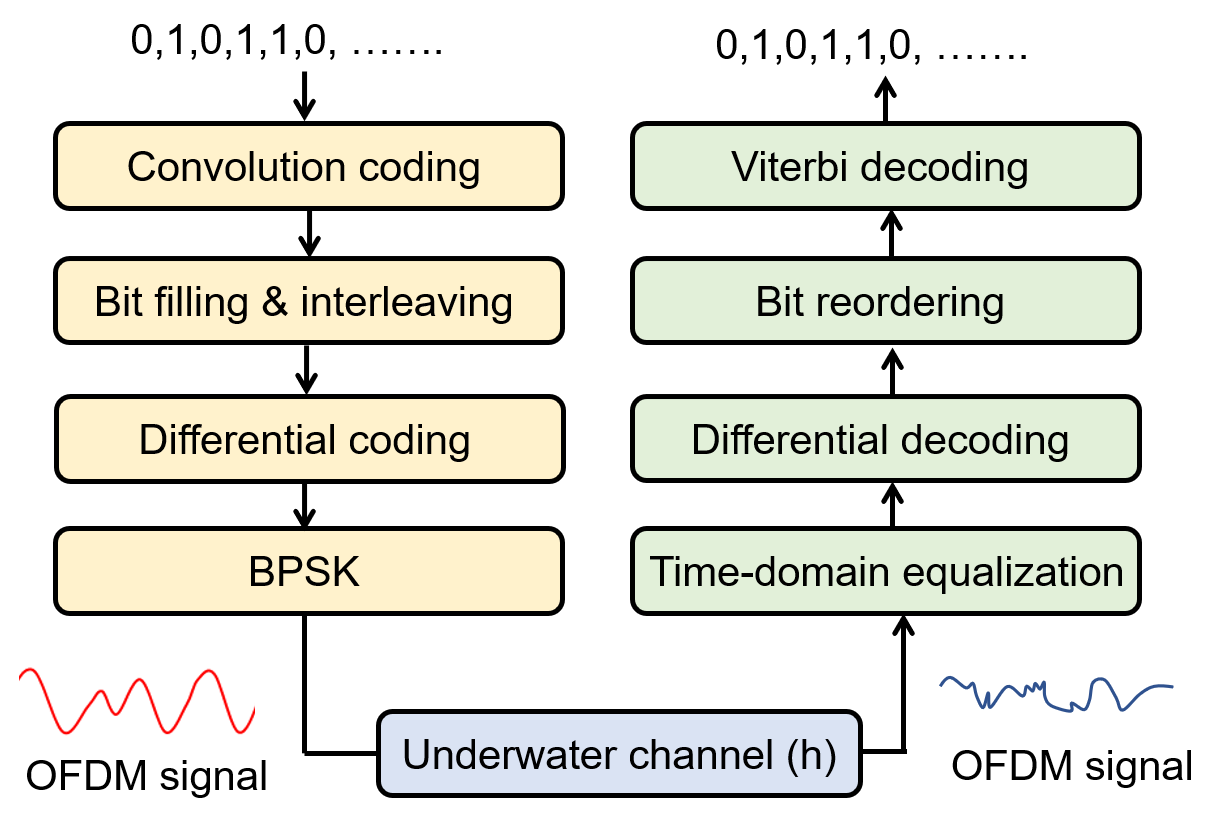}
    \vskip -0.1in
\caption{{\bf Data encoding and decoding.}}
\vskip -0.15in
\label{fig:data}
\end{figure}

\subsubsection{Encoding feedback}
We encode the frequency bins $f_{begin}$ and $f_{end}$, in a single OFDM symbol by assigning all the power only to  the  two corresponding bins.  By allocating all the power from the transmitter into two frequency bins, Alice can still decode the feedback signal reliably even when channel estimates of the backward path from Bob to Alice are unknown. Specifically, Alice can decode the signal by extracting the frequencies with the top-2 SNRs. The OFDM symbol from Bob would arrive at Alice after a delay of around a round-trip time. Since this specific OFDM symbol  is effectively the same as transmitting two frequency tones, Alice  performs an FFT over a sliding window with the same length as the OFDM symbol. The sliding window computation is performed up to  the maximum round trip time corresponding to 30~m, to search for Bob's OFDM symbol.

\subsection{Data transmission}\label{sec:data}
Multi-path  can be  severe under water due to reflections from the surface, floor  and other  objects~\cite{stojanovic2008underwater} causing  inter-symbol interference (ISI). To address  ISI without sacrificing  bit  rate, we apply  time-domain MMSE equalization  instead of  increasing the cyclic prefix. 
Dynamic channel changes due to motion of humans, waves and other underwater objects  is another challenge. Even within a packet, the channel for the  first OFDM symbol may differ from the last symbol, which may deteriorate the equalizer's performance.  Here, describe our  encoding and decoding mechanisms to address these issues.
   
%and the traditional OFDM system usually use longer 
%The max delay time underwater is usually . To avoid the 

%After the bandwidth adaptation, Alice transmits the data symbol which only encoded data in the selected frequency bins, and Bob needs to decode the data from the received signal. 
%Due to the complexity of real-world underwater environment, there are several challenges which need to be addressed to guarantee reliable data transmission. (1) Due to the 

\subsubsection{Data encoding.} 
 Alice  encodes the transmitted bits and generates  OFDM symbols  between $f_{begin}$ and $f_{end}$. The duration of each OFDM symbol is  960 samples (20 ms duration and 50~Hz subcarrier spacing with 48000 kHz sampling rate). We also add a  67 sample  cyclic prefix ($6.9\%$ overhead).  
%The encoding scheme consists of  convolution coding, bit filling, and differential coding. 

\squishlist
\item \textit{Convolutional coding.} We first apply  2/3 convolutional  coding with constraint length $K=7$ to encode  data bits. These codes are  widely used in  GSM and satellite networks~\cite{halonen2004gsm, butman1981performance}.
\item \textit{Interleaving bits.} 
The next step is to assign the coded bits  to OFDM symbols. Our  algorithm   has two main rules:  (1) assign bits to  OFDM bins between $f_{begin}$ and $f_{end}$ and set the remaining bins to zeros and  
(2) interleave the bits to avoid  consecutive bit errors which  could cause unresolved  errors after  convolutional decoding. Our  empirical observations show that   bit errors usually happen in the OFDM packets at a specific subcarrier or nearby two subcarriers. Our interleaving strategy is to  first fill one symbol until all selected subcarriers in this symbol have bits, and then fill the next symbol to avoid consecutive bit errors in a single subcarrier. Within a symbol,  after assign a bit in a  subcarrier, we leave some subcarriers for the  next bit with a step size is  one-third of the selected bins. If we use less than three  bins then this  defaults to not using  interleaving.    
%So during  interleaving, we avoid placing consecutive encoded bits in the same  subcarrier.

\item\textit{Differential coding.}  In our system, we use differential encoding across consecutive OFDM symbols to  alleviate the time-varying channel effects. Specifically, say  the data in  subcarrier $k$ of symbol $i-1$ is $y_{i-1}(k)$ and $b$ is a coded bit  intended for transmission. Then, we  apply  differential coding by setting the data in subcarrier $k$ of the next symbol $i$ as,  $y_i(k) = y_{i-1}(k)\bigoplus b$. Thus, the bit  is encoded as an XOR  between two consecutive OFDM  symbols. Since  channel variations across  two consecutive symbols is  small,  differential coding provides resilience to  channel variations, as long as the  coherence time is larger than one OFDM symbol.
\squishend
   
%Specifically, we assume the $y_{i-1}$ is the bit just transmitted and $x_{i}$ is a bit intended for transmission. So the next actual transmitted bit is the $y_i = y_{i-1}\bigoplus x_i$.
%After the bandwidth adaptation, Alice transmits the data symbol which only encoded data in the selected frequency bins. The encoding scheme consists of three phases: 
After assigning the  bits to the   OFDM subcarriers, we demodulate the bits in each subcarrier using BPSK and use IFFTs to generate the  time-domain OFDM signal. 

\subsubsection{Data decoding.} 
To decode data, Bob  first applies a 128 order FIR bandpass filter, with a  passband of 1--4 kHz, on the received signal to filter out  ambient  noise. %Our  decoder  consists of three main steps: MMSE equalization,  differential decoding, and   convolution decoding.  

\squishlist
\item\textit{MMSE Equalization}.
Time domain equalization  utilizes the equalizer coefficients to recover the transmitted signal and address inter-symbol interference (ISI). The equalizer coefficients is estimated using preset training symbols and the MMSE algorithm. The  communication channel model can be written as $y = h * x + n$, where $x$ is the transmitted  signal, $y$ is the received signal, $h$ is the channel coefficient vector with  length $L$, and $*$ is the convolution operator.  
The time-domain equalization model can be written as,  
$\hat{x} = g * y$, where $\hat{x}$ is the recovered signal, $y$ is the received signal, and $g$ is the equalizer coefficient with length $L$. MMSE equalizers~\cite{equalizer}  minimize the mean square error between the transmitted signal and the recovered  signal, i.e. $\min \|\hat{x} - x\|_2^2$. {In our system, a known training symbol is appended to the front of the data symbols. We utilize this training symbol to estimate our MMSE equalizer with the channel length $L$ of 480 samples. %To reduce computation, we calculate the estimated equalizer for each training symbol and average them to approximate the optimal equalizer. 
Finally, we apply the estimated equalizer to each received OFDM data symbol for transmitted signal recovery.}

\item\textit{Differential \& convolutional decoding.}  After   equalization, we  apply an FFT on each symbol and acquire the data in the  frequency bins. We  calculate the phase difference of two consecutive symbols on each OFDM bin within the frequency range of our  frequency band  adaptation algorithm. By measuring the phase difference, we can extract the coded bits, 0 or 1. 
We then re-interleave the bits following the order of the pre-determined interleaver. Finally, we use the Viterbi  algorithm with  constraint length  7. This provides maximum likelihood values for the data bits. % and further is  parallelizable for real-time implementation.
\squishend

We note the following points about our design.
\squishlist
\item {\it Doppler shifts.} A concern is that motion could also cause Doppler shifts, leading to  inter-channel interference (ICI)  within the OFDM symbol. Since our system mainly focuses on  underwater human communication, the safe motion speed for humans during scuba diving is usually slower than  1 m/s~\cite{wojtkow2017biomechanics}. Even if  the transmitter and receiver are moving in opposite directions, their relative speed is 2 m/s. The typical underwater acoustic speed is 1500 m/s. This results in a Doppler shift  of around 5~Hz at our maximum operational frequency. We however note that our OFDM subcarrier spacing is 50Hz, which is much higher than this Doppler shift and thus does not cause significant ICI. 
\item {\it Encoding ID and ACKs.} 
We use a single frequency in the OFDM symbol  to encode device ID as well as the ACKs. Specifically, ACKs are encoded by assigning the OFDM bin corresponding to 1~kHz to denote successful packet reception. This mechanism is reliable and does not require a long preamble since it assigns all the transmit power to a single OFDM bin. We also encode the device ID using the different subcarriers in the OFDM symbol. We have a total of 60 subcarriers and when the device transmits the OFDM symbol encoding its ID, it allocated all the power to the corresponding subcarrier. This limits the number of devices in our local network to 60 users, which may be  acceptable for underwater human activities like scuba diving and snorkeling. These IDs can be assigned using the app across devices in-air.

%%MOving SoS to implementation.
\squishend

\begin{figure*}[t!]
    \includegraphics[width=.16\textwidth]{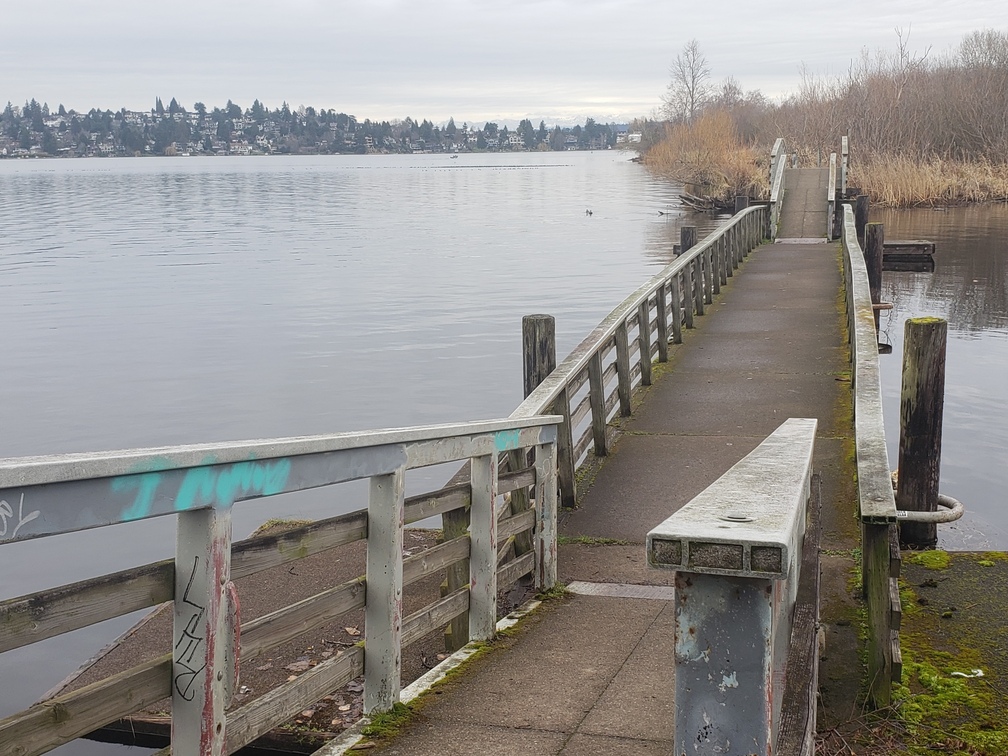}
    \includegraphics[width=.16\textwidth]{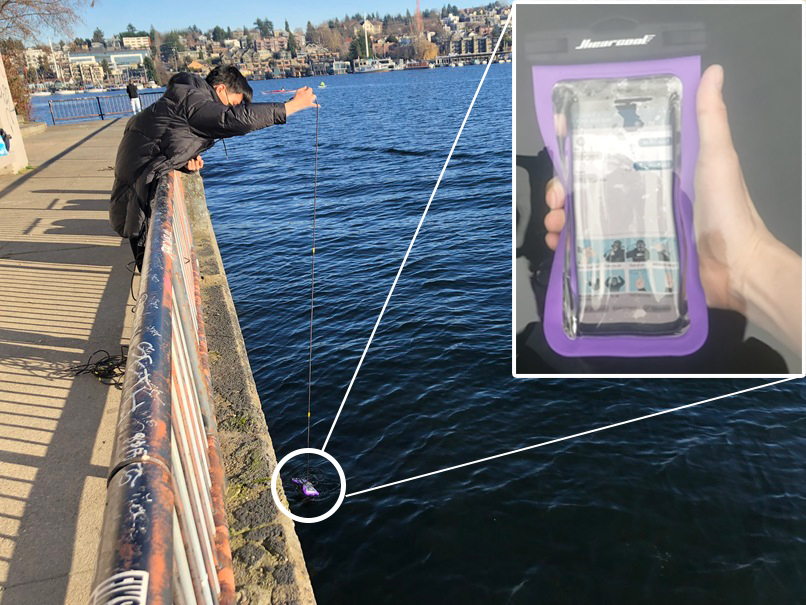}
    \includegraphics[width=.16\textwidth]{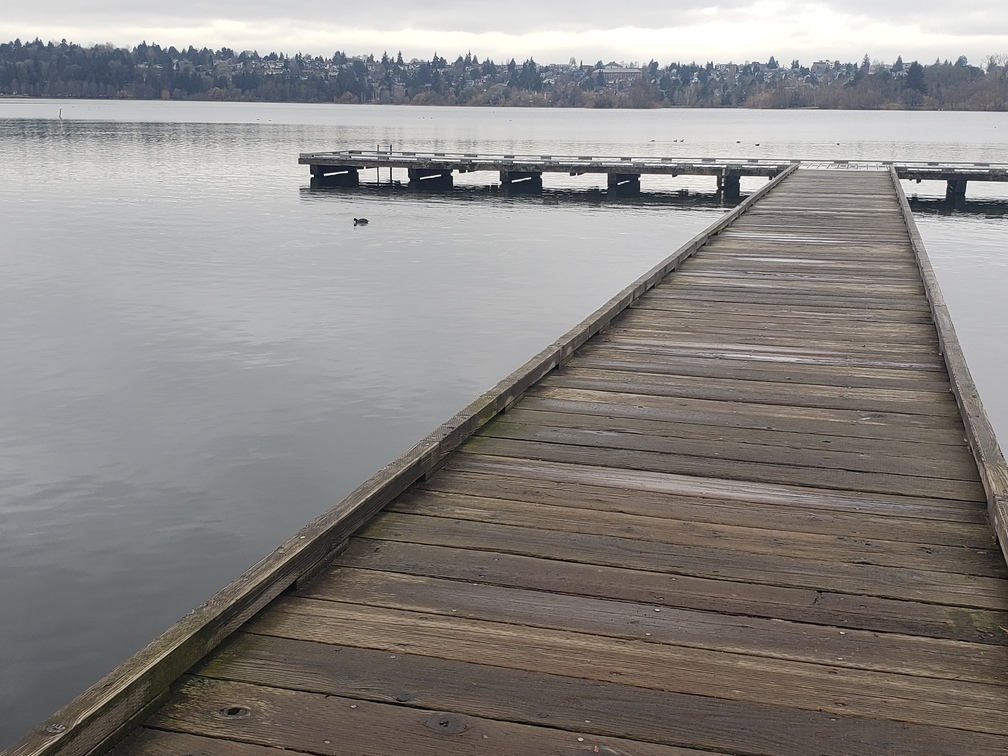}
    \includegraphics[width=.16\textwidth]{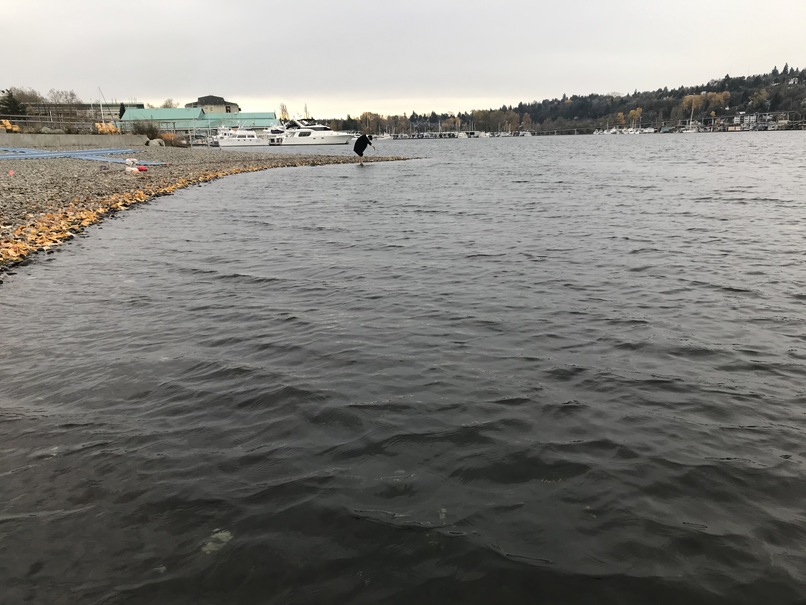}
    \includegraphics[width=.16\textwidth]{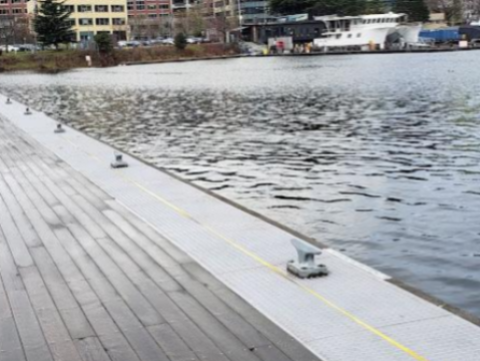}
    \includegraphics[width=.16\textwidth]{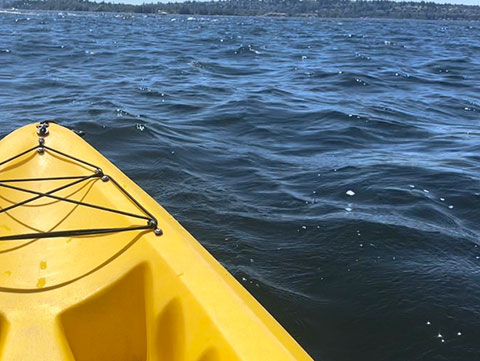}
    \vskip -0.15in
\caption{Different experimental environments. (a) Bridge: a quiet  environment. (b) Park: A busy location with boats and kayaks passing by regularly. (c) Lake: a busy and noisy location by a fishing dock with animals and other water activity. (d) Beach: a waterfront with a length  around 100~m for long range experiments. {(e) Museum: a highly occupied location used to dock ships and boats with a depth of 9~m for depth experiments. {(f) Bay: a location with 15~m depth  for deeper water experiments. }}}
\vskip -0.15in
\label{fig:locs}
\end{figure*}

\subsection{MAC protocol}\label{sec:mac}

In a typical use case, we expect our network to be operating in a low load scenario where not all transmitters send data at the same time. However, to support multiple devices that may operate at the same time, we also use carrier sense, similar to Wi-Fi, as a means of  mediating access to the channel between multiple transmitters. We implement our carrier sense in realtime using energy detection by measuring the average energy level in the 1--4~kHz band, which is the frequency range used for communication in our system. We perform this measurement every 80~ms. Prior to transmission, each phone measures if the energy level on the channel exceeds a predefined threshold. If it detects that the energy level exceeds the threshold, the phone waits for a random backoff period that is defined in multiples of the packet duration. During this backoff period, the phone continues to listens to the channel. If it detects energy on the channel during this backoff period, it will increase the backoff time by the duration of one packet to ensure the backoff period will not elapse while a packet is being transmitted on the channel. After this additional time has elapsed, the phone will again check  if the energy on the channel is below the threshold. After the remainder of the backoff time has elapsed and the channel is idle, the phone is then clear to send a packet. The threshold is computed by measuring the average noise level for a few seconds in each environment before use.

We note two points: 1) In addition to energy detection,  Wi-Fi receivers  also use preamble detection as part of carrier sense, which we could also incorporate  to improve noise resilience. 2) Our post-preamble feedback mechanism   can be considered as a light-weight version of RTS-CTS. So we can add a preamble to Bob's feedback which could be used in lieu of  a CTS message to address the  hidden terminal problem. Our current implementation does not include either features.

\section{Evaluation}

% \begin{table}
% \centering
% {\footnotesize
% \begin{tabular}{|c|c|c|}
% \hline
% {\bf Operation} & {\bf Smartphone (ms)} & {\bf Smart watch (ms)}\\ \hline
% Preamble detection &128 &xxx \\ \hline
% \makecell{Channel estimation\\Frequency adaptation} & 2&xxx \\ \hline
% Feedback decoding &1 &xxx \\ \hline
% \makecell{Packet decoding\\(24 symbols)}& 34 &xxx \\ \hline
% \end{tabular}
% }
% \caption{Runtime of all operations in our communication pipeline.}
% \label{table:runtime}
% \end{table}

 % We use the fftw3 library to perform FFT operations on the devices. In our implementation, each smart device is set to continuously listen for preambles on the channel  and running our preamble detection algorithm. %A device that is set to transmit a message is configured to transmit a preamble every 2~s until it receives a feedback signal from the receiving device. If it does not receive a feedback signal after 30~s, it will timeout.

 %{The minimum audio sample buffer size returned by the Android API is 80~ms. In our implementation, we maintain a FIFO history buffer of up to 240~ms, which is slightly  longer than the length of our  preamble. We are able to continuously run our preamble detection algorithm on this  buffer.  The runtime to execute our channel estimation, frequency adaptation, and feedback decoding algorithm are each on average 1--2~ms. Our decoder can perform equalization and Viterbi decoding for each symbol in less than 20~ms, which is the duration for an OFDM symbol.}
Our communication system has been implemented to run  on Android devices with preamble detection running continuously in real-time. The runtime to execute our channel estimation, frequency adaptation, and feedback decoding algorithm are each on average 1--2~ms on a Galaxy S9. Our decoder can perform equalization and Viterbi decoding for each symbol in less than 20~ms, which is the duration for an OFDM symbol. Fig.~\ref{fig:ui} shows our app interface where  users can select from a  list of 240 messages corresponding to hand signals used by professional divers. Users can filter the list of messages based on eight different categories. Additionally, the 20 most commonly used hand signals are displayed more prominently for selection by users. Pictorial representations of hand signals are included for reference in the app.  The size of our data packet is 16 bits, 24 bits after applying a 2/3 convolutional code. With this, users can  choose to send two hand signals in a single packet.

Our system also encodes a user's  6-bit ID into an SoS beacon using frequency-shift keying. Specifically, we encode a 0 bit with a single frequency tone $f_0$ and a 1 bit with a single frequency tone $f_1$. We design our system to support data rates of 5, 10, and 20~bps and use frequencies in  1.5--4~kHz  to transmit these  beacons over longer ranges. Using this  scheme we may also encode an 8-bit hand signal, which can be transmitted in around a second at these  rates.

We evaluated our  system in four  underwater environments  with different multipath effects and noise levels (Fig~\ref{fig:locs}). 

\squishlist
\item \textit{Bridge.} Under the water of a bridge with a horizontal distance of 20~m. This is a quiet  location with still waters.
\item \textit{Park.} By the waterfront of a park with a  length of 40~m. This is a busy location with boats and strong  currents. 
\item \textit{Lake.} Fishing dock by lake with a 30~m length. The lake had a 5~m depth. This is a busy location with people fishing and kayaking.
\item \textit{Beach.} The length of the water here is around  100 m. 
{\item \textit{Museum.} This location has a  depth of 9~m. This is a highly occupied location used as a dock for different boats and ships.}
{{\item \textit{Bay.} This location has a  depth of 15~m. There were a lot of waves at this location and experiments were performed on a kayak.}}
\squishend

\begin{figure}[t!]
\vskip -0.05in
    \includegraphics[width=.35\textwidth]{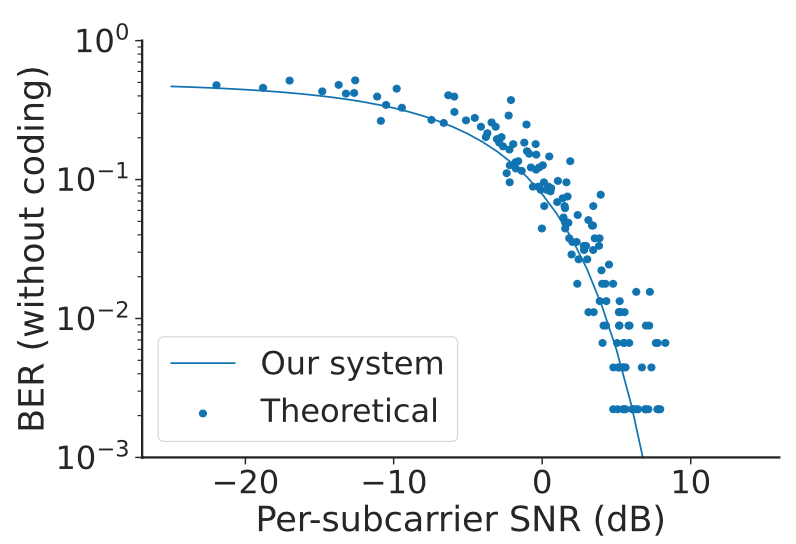}
\vskip -0.15in
\caption{BER measurements are from  distances of 5, 10, and 20~m with bandwidth of 1-4 kHz.}
\vskip -0.2in
\label{fig:snr_ber}
\end{figure}

\vskip 0.05in\noindent{\bf BER versus SNR.} First, we evaluate our decoding algorithm by computing the BER at different subcarrier SNRs. Given the frequency fading nature of underwater links, BER can be different at each of the OFDM subcarriers depending on its SNR. In these experiments we use our real-time implementation on two 
 Samsung Galaxy S9 phones. In all experiments, the phone speaker is set to its maximum volume. The phones were placed in a waterproof case and submerged at a depth of 1~m using a selfie stick as an extension pole. We perform the experiments at a distance of 5, 10, and 20~m at the bridge location. At each distance, we configure the transmitter to send a total of 500 OFDM symbols on the subcarriers between the 1--4~kHz and the modulation scheme for each subcarrier is set to BPSK.  We computed the BER as the fraction of mistakenly decoded bits (without coding) to the total transmitted bits over 500 symbols.  Fig.~\ref{fig:snr_ber} plots the BER curve obtained in our experiments in comparison to the theoretical curve for BPSK. The plot shows that the empirical data for our design  follows a similar trend to the theoretical estimates. %In particular we note that the SNR required to obtain a BER of 1\% in our system is on average 4~dB. %The signal power is the product of sending signal and the estimated channel, while the noise power is the differencebetween received signals and the product of sending signaland the estimated channel

%signal power as the squared channel estimate, and computed the noise power as the squared difference between the received signal and the transmitted signal multiplied by the channel estimate.

\begin{figure*}[t!]
    \includegraphics[width=.8\textwidth]{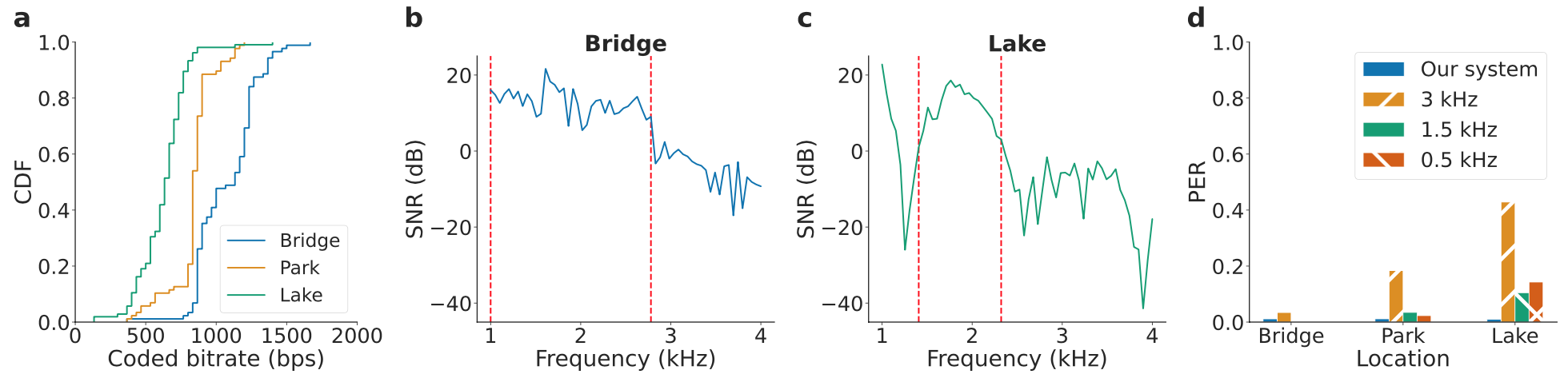}
    \vskip -0.05in
\caption{Effect of different environments. (a) CDF of bitrates selected by our algorithm. (b,c) Example frequency responses from two locations. Dashed lines indicate the range of selected frequencies by our algorithm.  (d) A PER value computed using  all packets at each location for our system and fixed bandwidth schemes.}
\vskip -0.05in
\label{fig:locs2}
\end{figure*}

\vskip 0.05in\noindent{\bf Effect of different environments.} We evaluate our system  in three locations: the bridge, park, and lake. These  range from quiet and still to noisy and busy, and capture a diversity of environmental effects. We report the  bitrate picked by our real-time algorithm and the PER of our system at  5~m. 

 During the experiment, our system runs in real-time where the transmitting smartphone first sends the preamble and header to the receiver. The receiver continuously listens for the preamble, when it  detects it,  performs SNR estimation on each OFDM subcarrier, runs our frequency band  adaptation algorithm and sends back  $f_{begin}$ and $f_{end}$ frequencies to the transmitting phone in a single OFDM symbol. The transmitting phone then sends data using OFDM  subcarriers between $f_{begin}$ and $f_{end}$. Each data packet  contains 2 bytes of  data (24 bits after applying a 2/3 convolutional code) on the selected OFDM frequency bins. This is more than sufficient to be able to encode the range of hand signals used by professional divers. We compare the performance of our frequency adaptation scheme to fixed-rate bandwidth schemes, where the transmitting phone sends the same bits using a fixed bandwidth of 1--4~kHz, 1--2.5~kHz and 1--1.5~kHz, which correspond to 60, 30, and 10 OFDM  bins respectively. This procedure was set to repeat 100 times at each of the three locations tested. Every 25 packets, we would pause measurements on the smartphone, retrieve it from the water, and submerge it again, until all 100 packets were transmitted. Even if  one bit error occurs at the decoder output, we mark it  as an erroneous packet. 
 Fig.~\ref{fig:locs2}a shows  that the selected bit rate varies  across location as well as runs. Within the same location, the selected bitrate  changes as the multipath   changes with time. The average selected bitrate was highest at the bridge location which is likely because the environmental noise and water currents are lowest in this setting. In comparison, at the park and lake environments, people fishing and kayaking nearby may have affected the channel, resulting in a selection of lower bitrates.  Fig.~\ref{fig:locs2}b,c shows an example received frequency spectrum in the bridge and lake locations. The dashed lines  indicate the range of frequencies selected by our algorithm for sending data.  In contrast to the bridge location,  the spectrum at the  lake location exhibit more frequent and deeper dips. In the lake  location, the wall and pillars underwater would reflect the acoustic signals resulting in more frequency selectivity.

Fig.~\ref{fig:locs2}d plots the PER  obtained for our system, as well as the three fixed bandwidth schemes at the three environments. The plot shows that the PER for the fixed bandwidth schemes increases in response to larger multipath in the environment. The PER is larger in the lake environment which exhibits the highest variability in the received spectrum. In contrast, the PER of our system remains low across all three locations with an average value of 1\%. The PER obtained by our system is also lower than those obtained by the fixed bandwidth schemes at the park and lake location. %This suggests that our system can adapt to frequency selective channels and can achieve a better PER than fixed bandwidth schemes.

\begin{figure}[t!]
\vskip -0.15in
    \includegraphics[width=.4\textwidth]{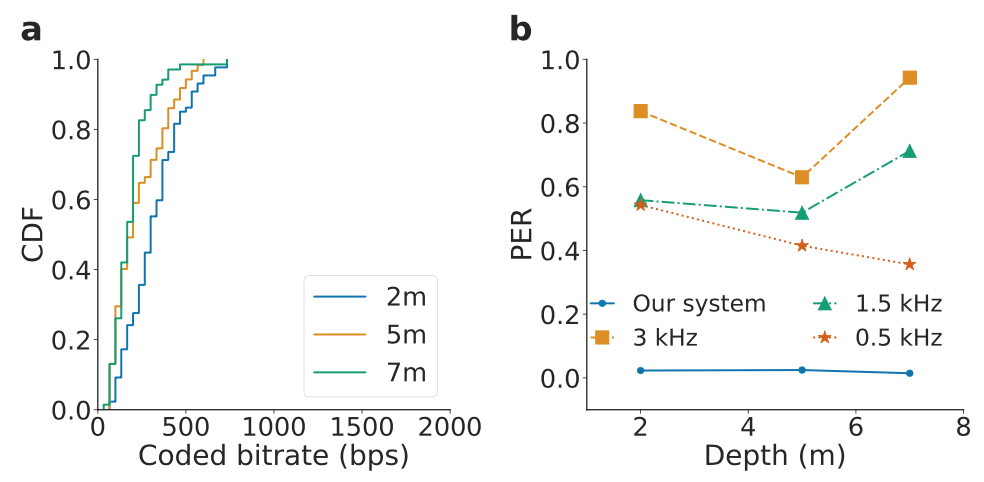}
    \vskip -0.15in
\caption{{{\bf Effect of depth. (a) CDF of selected coded bitrate. (b) PER our system  compared to fixed bandwidth schemes. A PER value is computed at each depth location using all packets sent at the location.}}}
\vskip -0.25in
\label{fig:depth}
\end{figure}

\vskip 0.05in\noindent{\bf Effect of different depths.} 
Multipath can change at different depths --- close to the surface, multipath interference from the surface of the water can be strong. Similarly, close to the bottom of the water body, it may experience significant multipath  from the floor. We perform our experiments in the {museum} location, which had a total depth of {9~m}. Our experiments were performed at a fixed horizontal distance of 5~m between the two smartphones. %We repeat the same experimental procedure as before,  when the phones were submerged to a depth of 1, 2, and 4~m. 

Fig.~\ref{fig:depth}a shows the bitrates selected by our system for three depths of {2, 5 and 7~m} and Fig.~\ref{fig:depth}b shows the PER. {The PER for our system and the 0.5~kHz fixed bandwidth scheme was highest at a depth of 2~m when the phones are close to the water surface. For the 1.5 and 3~kHz fixed bandwidth scheme, the PER was highest at a depth of 7~m when the phones are close to the bottom of the lake. These results suggest that the environments at a depth of 2 and 7~m are the most challenging multipath environments. This is likely because there are more objects for the signal to reflect from at the surface and bottom of the lake including ducks, fish, and kayaks. At all depths, our system obtained significantly lower PERs than the fixed bandwidth schemes.}

% The PER is highest at a depth of 1~m when the phones are close to the water surface. The PER is lower at a depth of 2~m, and rises slightly at 4~m when the phones are close to the bottom of the lake. These results suggest that the environment at a depth of 1~m is the most challenging multipath environment. This is likely because there are more objects for the signal to reflect from at the surface of the lake including ducks, fish, and kayaks. At all depths, our system obtained significantly lower PERs than the fixed bandwidth schemes.

\begin{figure}[t!]
\vskip 0.2in
    \includegraphics[width=.48\textwidth]{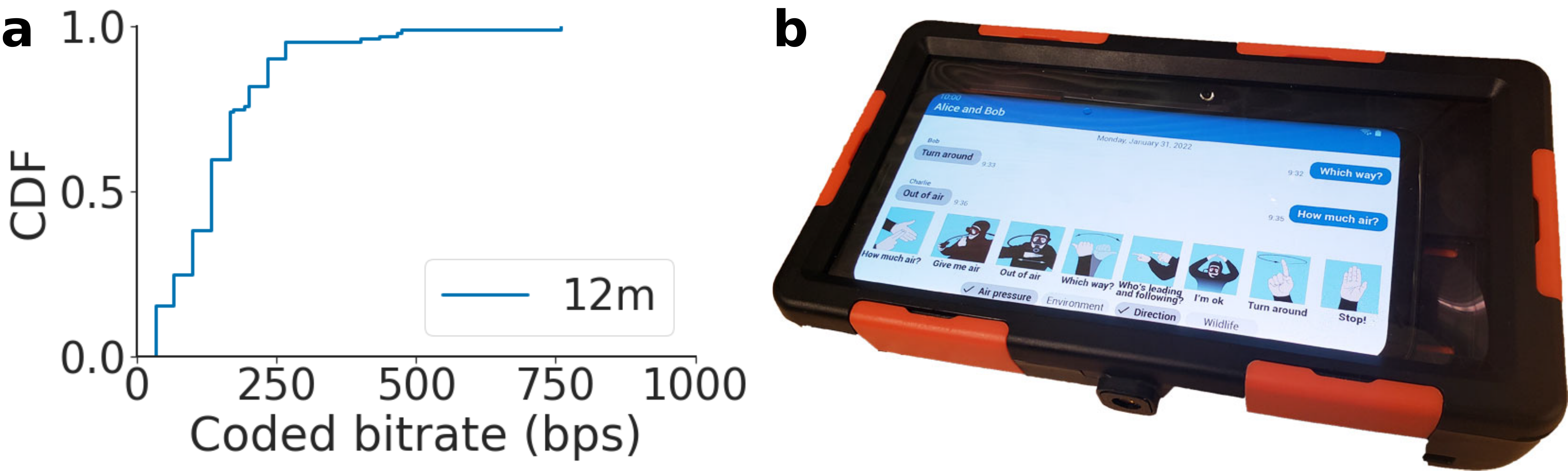}
    \vskip -0.15in
\caption{{Testing in deeper waters. (a) CDF of selected coded bitrates. (b) Waterproof casing rated for a depth of 15~m.}}
\vskip -0.15in
\label{fig:deeper}
\end{figure}

\begin{figure*}[t!]
    \includegraphics[width=\textwidth]{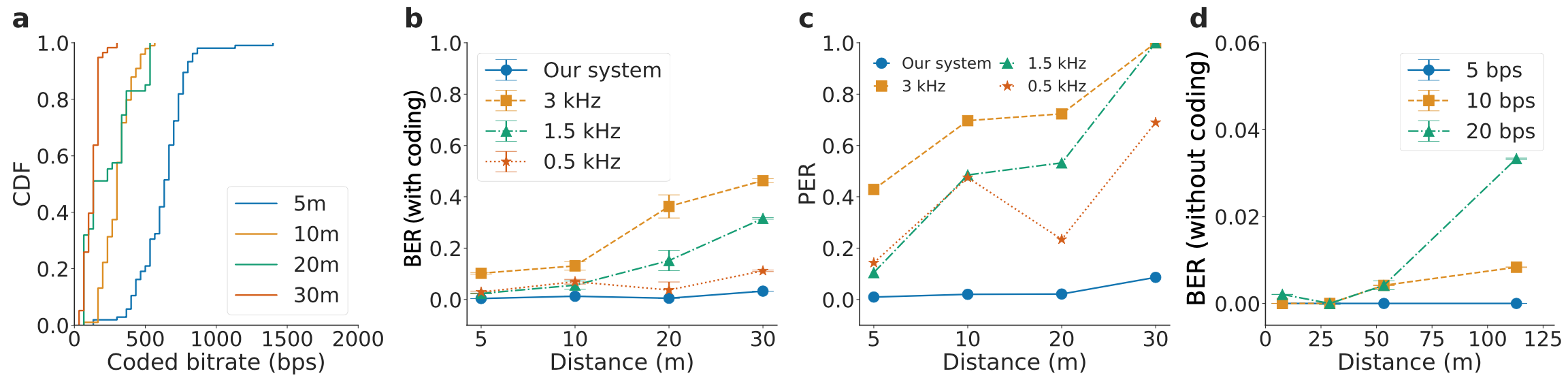}
    \vskip -0.15in
\caption{Range evaluation. (a) CDF of selected bitrates. (b,c) BER and PER of our system  versus fixed bandwidth schemes. A PER value is computed from all packets at the location.   (d) Longer  ranges using lower bit rates.}
\vskip -0.15in
\label{fig:distance}
\end{figure*}

\begin{figure}[t!]
    \includegraphics[width=.46\textwidth]{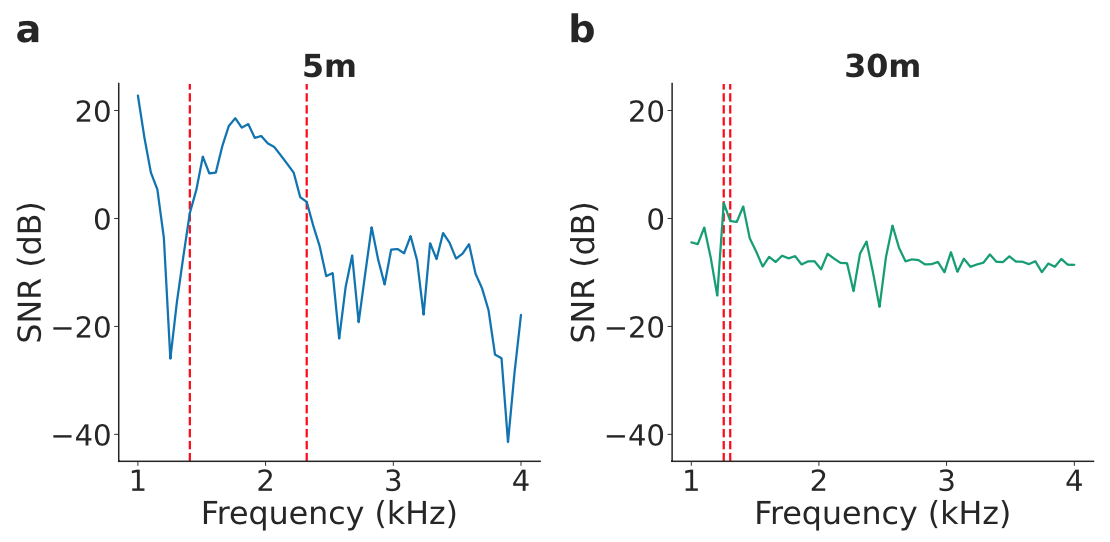}
    \vskip -0.15in
\caption{(a,b) Our system  uses a smaller band  in response to increased attenuation  at larger distances. }
\vskip -0.15in
\label{fig:dist_spec}
\end{figure}

\vskip 0.05in\noindent{\bf Testing in deeper waters.} {To evaluate our system at deeper depths, we performed an experiment at the bay location with a total depth of 15~m and submerged the phones to a depth of around 12~m. We note that 12~m is in the range of depths associated with basic recreational scuba dives~\cite{depth}. In this experiment, we used a different waterproof pouch~\cite{case2} (Fig.~\ref{fig:deeper}b) which was rated to work at a maximum depth of 15~m. We note that this pouch was made of a hard plastic casing (polycarbonate and thermoplastic polyurethane) which attenuated the sound more than the thin flexible plastic material (polyvinyl chloride) used in the pouch for other experiments. We positioned the transmitter and receiver phone on either side of a two-person kayak (3.5~m length), and weighed down the phones using a pouch of sand to ensure the phones would submerge underwater. Fig.~\ref{fig:deeper}a shows that the median bitrate selected by our system at a depth of 12~m was 133~bps. This suggests the phones are able to communicate even at these deeper depths and with a hard plastic casing.}

\vskip 0.05in\noindent{\bf Range evaluation.} 
We perform our range evaluation in the  lake location  since it had a comparatively long distance of 30~m. We performed our measurements at a depth of 1~m, which as we note in the previous evaluation, has empirically  more  challenging multipath  at the lake location. The phones were submerged into the water using a rope, which caused the phone to sway and rotate slowly during  measurements, i.e., the phones were not static.

 Fig.~\ref{fig:distance}a shows the  bit rate after coding selected by our system at different distances. The plot shows that the selected bitrate generally decreases with  distance, with the largest drop  occurring between 5 and 10~m. At a distance of 5~m and 30~m, the median selected bitrate was 633.3 and 133.3~bps respectively. In Fig.~\ref{fig:distance}b we also show the BERs for the coded bits transmitted at different distances for our frequency adaptation scheme and the fixed bandwidth schemes.  The BERs for the fixed adaptation schemes increase quickly with distance. This is because the fixed schemes will continue to naively transmit bits on low SNR subcarriers and increase the likelihood of bit errors. This is clearer in Fig.~\ref{fig:distance}c where we plot the PERs for these schemes over distance.  The plots show that the PER for the fixed adaptation schemes reach 100\% when using a fixed bandwidth of  1.5 and 3~kHz. In contrast, our frequency adaptation scheme has a PER of 7\%  at a distance of 30~m. This demonstrates that to minimize PER it is essential to pick the appropriate frequency band since even with a fixed low bandwidth of 0.5~kHz, it is likely that some of the  frequencies in this narrowband signal are in a deep fade resulting in sustained  packet losses despite using coding.  In contrast, our real-time adaptive system picks a conservative set of frequencies depending on the frequency profile and SNRs as shown in Fig.~\ref{fig:dist_spec} which allows it to minimize the packet error rate. %Note that in the case of the example spectrum at 30~m, a bandwidth of xxx~Hz (xxx~bps) is selected. At this bitrate, the time to send a 32-bit payload  is xxx~ms, which we believe is acceptable which is more than sufficient to send the 200 hand signals.

In the above design, each OFDM symbol had a 20~ms duration, limiting the minimum symbol rate to 50~bps. To further reduce the bit rate, we increase the symbol duration to 50, 100 and 200~ms and use a single frequency within each symbol to encode data. This results in a bit rate of 20, 10 and 5~bps respectively, which may result in a longer range. To evaluate this, we perform our experiments in the beach location. Testing was performed at a fixed depth of 1~m.  We compute BER as the number of bits that were correctly decoded over all the bits transmitted at each location at each of the three bit rates.  Fig.~\ref{fig:distance}d shows the uncoded BER of our system up to a distance of 113~m. The plot shows that the BER is less than 1\% for bitrates of 5 and 10~bps up to the maximum tested distance of 113~m. This demonstrates that our system enables  communication between smartphones  underwater at long ranges albeit at lower bit rates. We also note that a bit rate of 10 bps is sufficient to transmit SoS beacons which would be important at these long ranges during underwater activities. Similarly transmitting a 8 bit packet  that is sufficient to encode the 200 hand signals can also be done in around a second after accounting for coding.

\begin{figure*}[t!]
    \includegraphics[width=.78\textwidth]{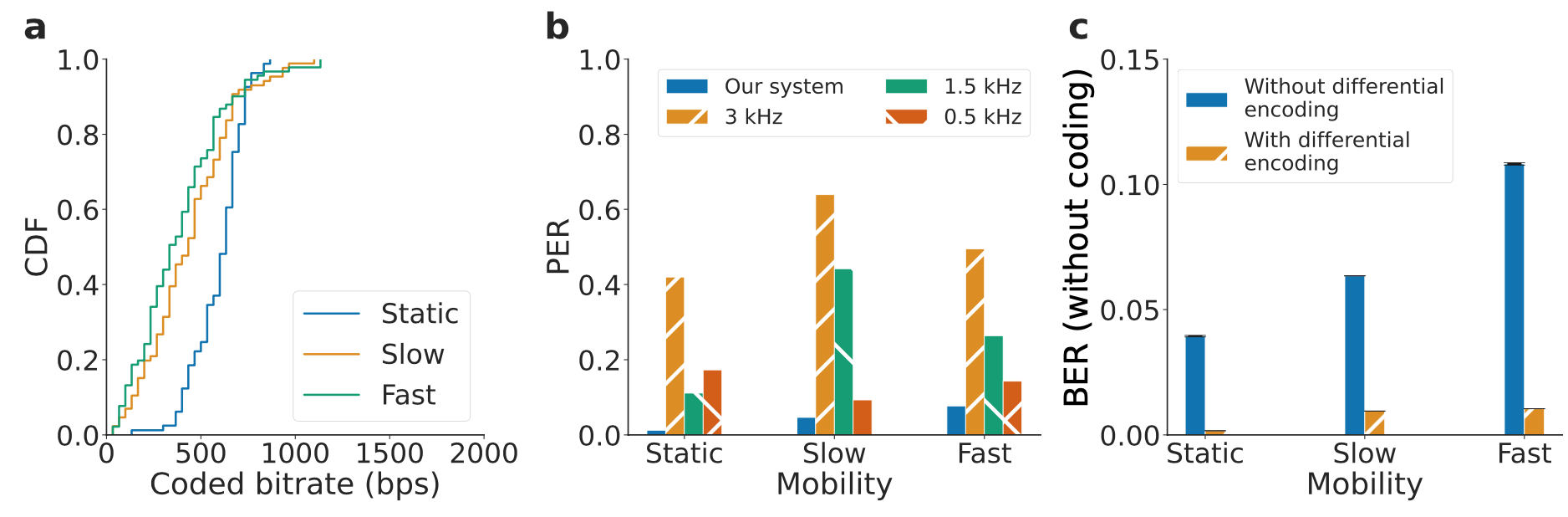}
    \vskip -0.15in
\caption{Effect of mobility. (a) CDF of selected bitrates in a static scenario versus one of the smartphones moving slow ($2.5 m/s^2$) and fast ($5.1 m/s^2$). (b,c) PER and uncoded BER of our system versus other configurations.}
\vskip -0.15in
\label{fig:mobility}
\end{figure*}

\vskip 0.05in\noindent{\bf Effect of mobility.}  We perform mobility evaluation in the lake location at a fixed horizontal distance of 5~m and at a depth of 1~m. In this experiment, we measure system performance when the phones are static, and when the transmitting phone is moving slowly and moving quickly. To do this, we  move the phone  horizontally back and forth, and vertically up and down. Since the phone is attached to the rope, the phones will also rotate randomly during the movement. During these measurements, the {raw} accelerometer {readings} on the  phone after compensating for gravity were on average, $2.5$ and $5.1 m/s^2$ for  slow and fast motion, respectively.
%Unlike previous experiments, we keep the phone submerged in the water for the entire duration of the measurement. In other words, it was not taken out of the water at any time. This ensures that changes to the multipath are largely due to moving the phones under the water. 

In Fig.~\ref{fig:mobility}c, we calculate the uncoded BER with and without differential coding. Specifically, for  received packets without differential coding, we only  apply the equalizer to each OFDM symbol and compare the decoded bits with the original bits.  The plot shows that  with mobility, the BER without differential coding increases rapidly and easily surpass $10\%$. When we apply  differential coding, the BER  decreases  dramatically and could still keep around $1\%$ in the fast motion scenario. Fig.~\ref{fig:mobility}a shows the CDF of the coded bitrate across these three mobility scenarios. We see that the coded bitrate is highest in the static situation with a median bitrate of 640~bps. In contrast, the median bitrates in the slow and fast moving case decrease to 433 and 336~bps respectively.  Fig.~\ref{fig:mobility}b shows that our PER   increases    from 1.2 to 7.6\% as mobility  increases.

\vskip 0.05in\noindent{\bf Effect of phone orientation.} 
 For this evaluation, we ran our system at the bridge environment with a fixed  distance of 5~m and a  depth of 1~m. The phones were submerged with the screen facing towards the water surface, and were first set so the speaker and microphones of each phone are directly facing each other.  We rotate one phone in the azimuth angle from 0 to 180° in increments of 45°. Fig.~\ref{fig:rot}a  shows that the median bitrate decreases with phone orientation from 1067~bps at 0° to 567~bps at 180°.  Fig.~\ref{fig:rot}b shows  that while the fixed bandwidth schemes  have a higher PER at large angles, our frequency adaptation scheme achieves a low PER. This is because our system is able to select a better  frequency band  and  adapt to the  channel at  different orientations.

{\vskip 0.05in\noindent{\bf Channel stability and SNR.} 
We take a closer look at the effects of mobility from the perspective of channel stability and SNR. Mobility has two effects on our system: (1) the channel may change between the preamble and the data symbols, leading to a different bandwidth selection and, (2) the channel between the first and the last data symbol may be different. For the second effect, as shown in Fig.~\ref{fig:mobility}(c), differential coding addresses the channel changes within the data packet. Here, we investigate the stability of the underwater channel between the preamble and data symbols in  different motion scenarios. In our system, Alice first transmits a preamble to Bob and Bob selects the bandwidth. To verify the channel stability and its effects on SNR, instead of transmitting the data symbols, we configure Alice to transmit another preamble; Bob used the bandwidth from the first preamble to compute the SNRs of the corresponding OFDM bins using the second preamble. 
Now, we can evaluate the effect of channel stability on our bandwidth selection algorithm. Specifically, when the SNRs in some of the subcarriers  within the selected bandwidth are very low,  packet errors can happen. Hence, we select the minimum SNR computed using the second preamble within the selected bandwidth as the metric to evaluate the performance of our bandwidth selection algorithm.
%Note that while conventional channel stability looks like variations in amplitude and phase in each OFDM bin, we do not use the amplitude and phase estimates from the preamble for decoding the data symbols. So we focus on bandwidth selection as a proxy for channel since it directly affects the PER in our system.
}

\begin{figure}[t!]
    \includegraphics[width=.46\textwidth]{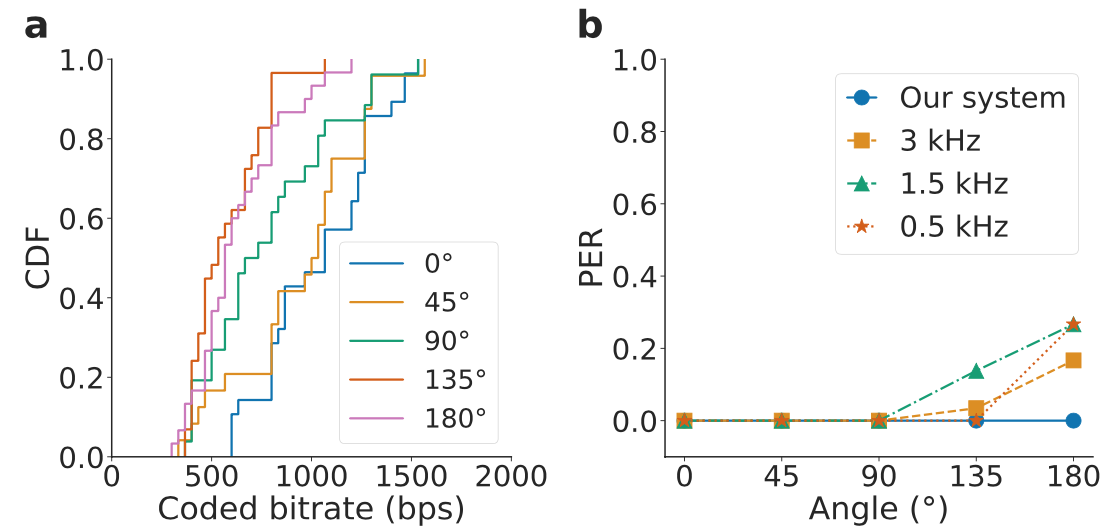}
% \vskip -0.2in
\caption{Effect of phone orientation. (a) CDF of selected bitrates for different azimuth  offsets between a pair of smartphones. (b) PER at a distance of 5~m for our system and other fixed frequency band schemes.}
\vskip -0.2in
\label{fig:rot}
\end{figure}

{
We performed this experiment at a horizontal distance of 10~m at the lake location when both phones were held static, and when moving them at slow and fast speeds similar to our mobility evaluations. The x-axis in 
Fig.~\ref{fig:channel}  represents the experiment index number, and the y-axis is the minimum SNR computed using the  second preamble within the selected bandwidth. The dashed SNR  line of 4~dB is the reference for subcarrier quality evaluation (4~dB may cause 1\% BER according to the SNR-BER curve). }

\begin{figure*}[t!]
    \includegraphics[width=.78\textwidth]{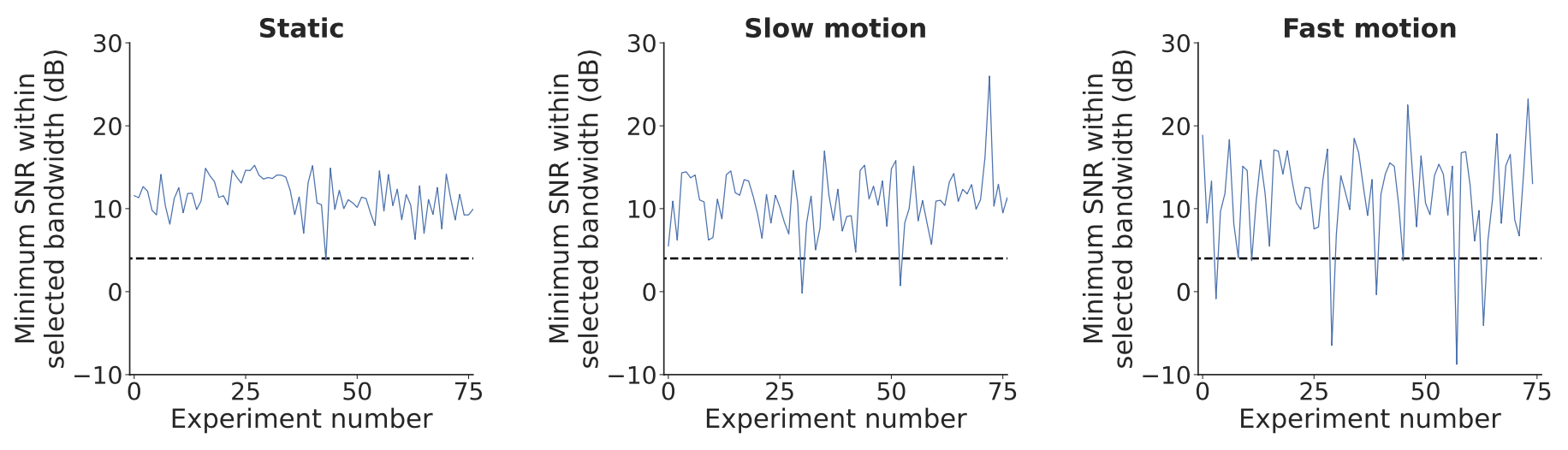}
    \vskip -0.15in
\caption{{{\bf Channel stability and SNR. For consecutively transmitted pairs of preamble signals separated by the feedback duration, the figure shows the minimum SNR computed using the second preamble over  the bandwidth picked from the first preamble. We run these experiments in  (a) static scenarios and with (b) slow and (c) fast motion. The dashed line at 4~dB marks the SNR threshold at which BER will theoretically be at 1\%, representing the  threshold for a stable communication link.}}}
\vskip -0.15in
\label{fig:channel}
\end{figure*}

\begin{figure*}[t!]
    \includegraphics[width=.78\textwidth]{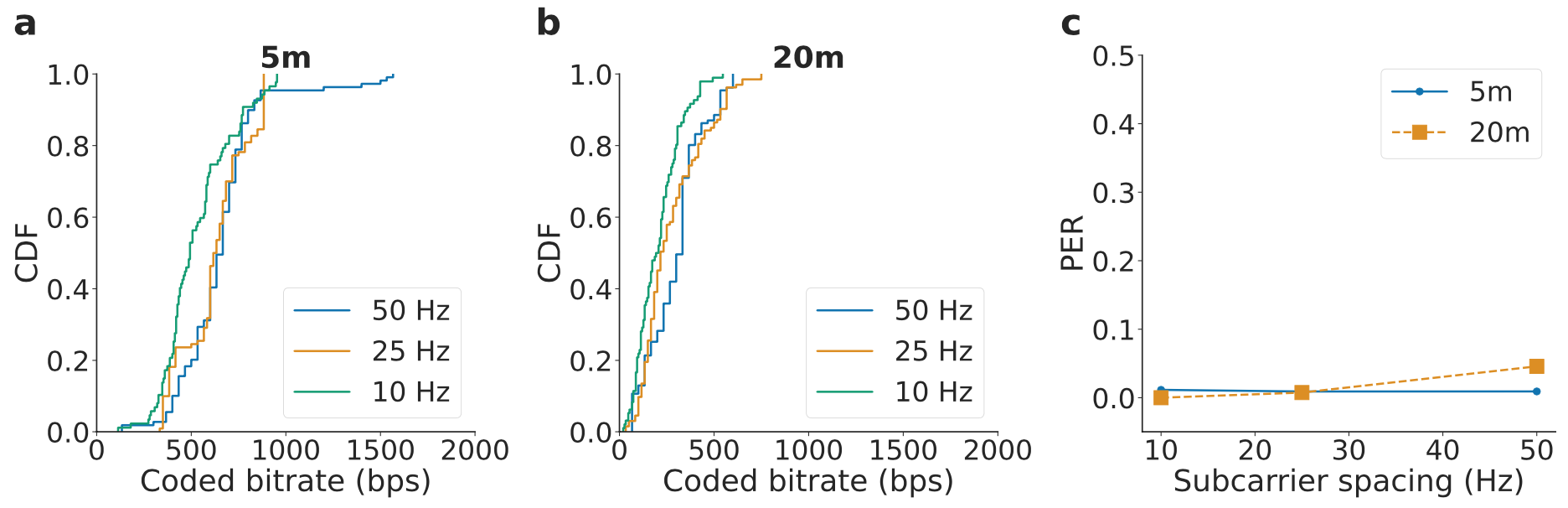}
\vskip -0.15in
\caption{{Effect of OFDM subcarrier spacing. CDF of selected bitrates for different subcarrier spacing values at a distance of (a) 5 and (b) 20~m. (c) PER for our system at different subcarrier spacings.}}
\vskip -0.15in
\label{fig:ns}
\end{figure*}

{In the static scenario,  the minimum SNRs are high, likely due to the conservative parameter settings used in our bandwidth adaptive algorithm. In the slow and fast motion scenario, the fluctuation of the minimum SNR values increase and sometimes the bad subcarriers appear in the bandwidth selection algorithm, leading to an  increasing PER during fast motion.
Despite the existence of channel variance with motion, two components in our system help  alleviate the effects of mobility  and keep the PER low: 
(1) the conservative parameter settings of our bandwidth adaptive algorithm with a  high SNR threshold and conservative factor $\lambda$ allows for some room for safe bandwidth selection albeit the inaccurate SNR estimation due to mobility. Due to these conservative settings, only a small proportion of packets would pick up the bad subcarrier (<4 dB). 
%Thus, as long as the difference in the bandwidth selection is not huge, we can still  minimize the PER.  
(2) In scenarios where our  conservative parameters can fail when the change of channel is very large, convolution coding helps reduce the bit errors. 
However, when channel drastically changes leading to more deep drops in the selected bandwidth, packet error can occur, which accounts for the increasing PER during fast motion. In future work,  further improvements can be used to make the bandwidth selection  algorithm to be even more conservative, which can lead to lower data rates. Another approach to improving the PER is to apply non-linear equalization techniques, which perform better for signal recovery in poor channels with severe distortion~\cite{paliwal2016comparison}. }

%the selected bandwidth from the first preamble and the y-axis represents the selected bandwidth using the second preamble. Note that in the figure we have also added some jitters to each point, to more easily visualize overlapped points. 
%In the static scenario, the mean absolute difference between selected bandwidths across two preamble is 61~Hz, which is approximately the size of one OFDM bin in our system. 
%In the slow and fast motion case, the mean absolute difference corresponds to approximately 136 and 169~Hz respectively. When the motion speed increases, the variance in the bandwidth selection increases. Therefore, the PER would increase when motion speed increases as confirmed  in  Fig.~\ref{fig:mobility}(b). 

%Such channel change may have two types of results. 
%Firstly, the selected bandwidth in the first preamble is lower than the selected bandwidth in the second preamble. Even if the data rate maybe sacrificed, the BER for data symbol can be still good due to the more conservative selection.

{
\vskip 0.05in\noindent{\bf Effect of OFDM subcarrier spacing.} 
Next, we perform experiments with three different OFDM subcarrier spacing: 50 Hz (length of OFDM symbol is 20~ms),  25 Hz (40~ms), and 10 Hz (100~ms). We do not select the OFDM subcarrier spacing lower than 10~Hz due to  Doppler shifts as described in~\xref{sec:data}. We perform these experiments in the lake location at a horizontal distance of 5 m and 20 m. During the experiment, for each subcarrier spacing, the OFDM symbol length in the preamble is kept the same as the length of data OFDM symbol to maintain the same frequency resolution. Fig.~\ref{fig:ns}(c) shows that at 5~m, all the PERs  are around 1\%. At 20~m, the PER with 50 Hz spacing increases to 4.6\%, while PERs for 25 Hz and 10 Hz spacings are lower than 1\%. Smaller subcarrier spacing can improve PER  because (1) smaller spacing can provide high-resolution SNR estimation and more accurate bandwidth selection and (2) smaller spacing can also improve the equalization performance due to the higher frequency-resolution in the training symbol. %On the other hand, in Fig.~\ref{fig:ns}(a,b), smaller subcarrier spacing does not improve data rate. The reason is that, even if the smaller subcarrier spacing can provide more accurate estimation of channel SNR, our system performs the bandwidth selection instead of subcarrier selection due to the limitation of feedback communication. In conclusion, smaller subcarrier spacing can indeed improve the SNR estimation of the channel and has lower PER. However, the smaller subcarrier spacing does not have higher data rate and it requires longer preamble for SNR estimation.
}

\vskip 0.05in\noindent{\bf Effect of air in water-proof case.}   To evaluate this, we first expelled as much air as we can from the water-proof case before putting the phones into the water and then measure  the frequency response. We then filled the case with air before putting the phones in the water and then analyzed the frequency response. Fig.~\ref{fig:air} shows that even if the frequency response of the two curves has some difference, the average power within 1-4~kHz  was not significantly different. %Therefore, a bit of air inside the case does not  significantly attenuate the signals.

%\vskip 0.05in\noindent{\bf Using two  microphones.} Next, we evaluate how dual microphones could improve  performance compared. We  use the collected data from the previous orientation experiment.  For the each received packet in different orientations, we apply three different decoding methods: (1) only use the signal from the bottom microphone, (2) only use the signal from the top microphone, (3) use   signals from both microphones and our combination algorithm for decoding. The uncoded BER of these three methods  is  shown  in Fig.~\ref{fig:multimic}.  From Fig.~\ref{fig:multimic}, we have several findings: (1) Our dual-microphone combination algorithm can help to improve the communications quality compare with single microphone. (2) At all rotations angles, the BER from top microphones is higher than the BER from bottom microphone. This phenomenon implies that the dual microphones on the phone is not indentical, because of their different utility. Another possible reason is that the structure of the waterproof case near the two microphones is different, which may lead to different acoustic response characteristics.

%\begin{figure}[t!]
 %   \includegraphics[width=.48\textwidth]{./figs/multimic2}
%\caption{Multiple microphones}
%\label{fig:multimic}
%\end{figure}

%\begin{figure}[t!]
 %   \includegraphics[width=.48\textwidth]{./figs/preamble}
%\caption{Performance of preamble detection. The figure shows that we can effectively correlate with the packet preamble across the full tested distance of 30~m.}
%\label{fig:preamble}
%\end{figure}

\vskip 0.05in\noindent{\bf Preamble \& feedback signal evaluation.} We also perform preamble evaluation in the lake location at a depth of 1~m. We  transmit 180 preambles at each distance and evaluated the probability that our system could successfully detect the preamble. Our measurements show that our detection rate defined as the fraction of detected and transmitted preambles was 0.99, 1.0, 1.0 and 0.96 at 5, 10, 20 and 30~m. 
We also measured our system's ability to correctly decode the feedback signal containing the result of our frequency adaptation algorithm at all the above distances.  Frequency error rate is computed as the fraction of feedback signals where the  decoded frequencies did not match the transmitted values. Across all tested distances, the error rate was around 0.01, i.e., 1 in 100 packets across all these distances. This is because we  allocate all the power to these two frequencies making the signal strength much higher which allows reliable decoding even with frequency fading.
In the cases where there are errors, the system confuses it to the adjacent OFDM  bins.   %This shows that  our system can reliably decode  feedback  using our encoding mechanism.

\vskip 0.05in\noindent{\bf MAC protocol evaluation.}    Finally,  we measure the effectiveness of carrier sense  at supporting multiple devices underwater.  We consider two network deployments with  four phones (three transmitters, and one receiver) and three phones (two transmitters, and one receiver) placed underwater in the bridge location at a depth of 1~m at distances in the range of 5-10~m from the receiver. As mentioned before, we expect that not all devices will be transmitting continuously underwater. But to test the effectiveness of carrier sense in a heavily used network, we  configure the transmitters to send continuously  after a random backoff period of multiple seconds, up to a maximum of 120 packets per transmitter. We repeat the experiments with and without carrier sense.

We measure the fraction of collisions for each of the transmitter. To calculate this fraction, we first logged the timestamps at when the transmitting phones sent a packet. Packets that were transmitted within one packet duration of each other was marked as a collision. The fraction of collisions in the network was defined as the number of packets that were involved in a collision divided by the total number of packets sent by all transmitters (dotted horizontal lines).  Fig.~\ref{fig:mac} shows that without carrier sense, the collision rate in the three transmitter  network is high with an average of 53\%, however, our carrier sense mechanism is able to reduce this to an average of 7\%. The figure shows a similar collision reduction  for the two transmitter  network from  33\%  to 5\%.

%Fig.~\ref{fig:fsk} shows the bit error rate as a function of distance. We observe that the error is higher at closer distances, we find that this is due to harmonics and intermodulation distortions that are more pronounced in the received signal at close distances, which causes false peaks to be registered by our peak detection algorithm. However, at a distance of 20 and 30~m, the plot shows an error rate of around than 1\%, i.e., 1 in 100 packets across all these distances.  This shows that  our system can correctly decode the feedback signal using our encoding mechanism.

%\begin{figure}[t!]
 %   \includegraphics[width=.48\textwidth]{./figs/fsk}
%\caption{Performance of feedback signal detection.}
%\label{fig:fsk}
%\end{figure}

% \begin{figure}[t!]
%     \includegraphics[width=.48\textwidth]{./figs/sos}
% \caption{Performance of SoS beacons. The plot shows that we can achieve long range communication of emergency SoS messages on smartphones underwater at different bitrates.}
% \label{fig:sos}
% \end{figure}

\begin{figure}[t!]
    \includegraphics[width=.4\textwidth]{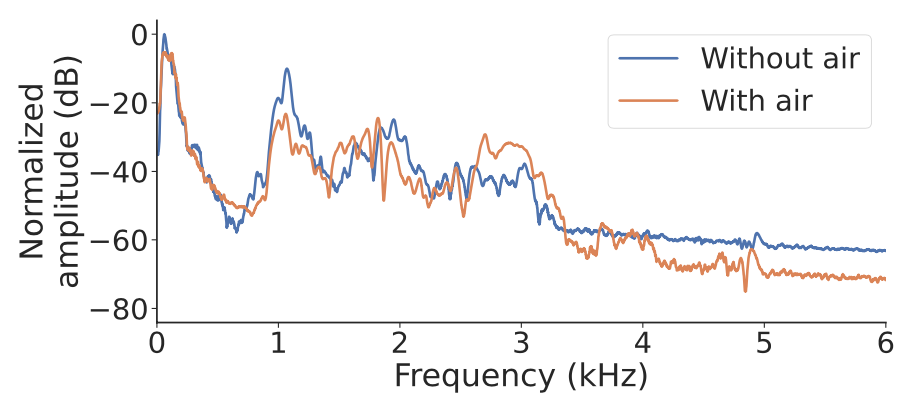}
    \vskip -0.15in
\caption{{\bf  Effect of air in waterproof case.}}
\vskip -0.25in
\label{fig:air}
\end{figure}

\section{Related work}
Underwater networking is an active research area where  topics across the network stack   continue to be  explored for custom acoustic modems and sensor networks~\cite{mac-survey}. There is also prior work in localization and ranging~\cite{loc2,loc1}, which is not in scope of our work.  Below we describe the prior work  in more detail.

\noindent{\bf Acoustic modem hardware.}  Since acoustic signals propagate well underwater, unlike RF, the  common  approach is to design custom acoustic modem hardware and use custom amplifiers at high power to achieve long ranges. Given the lack of economies of scale, much of this hardware can be expensive or not as accessible as in-air   radios. All prior works  that present underwater network  simulators~\cite{seanet,seanet2,unet} or  protocol stacks~\cite{netarch,sunset,desert} use  custom hardware with amplifiers and (de)modulators. Recent work integrates a custom acoustic modem with software radios  (e.g., USRPs) to design a software-defined platform for underwater networking research~\cite{cog-usrp,usrp2}. The closest to our work is  iSonar~\cite{isonar} that designs a custom acoustic OFDM  modem hardware that can connect to the smartphone using its audio jack. In addition to requiring additional hardware, it neither supports band adaptation nor is evaluated with mobility. In contrast, we design the first underwater acoustic communication system that operates on mobile devices without any additional hardware.

\noindent{\bf Modulation, rate adaptation and  MAC protocols.} Prior work  makes contributions at  the physical and MAC layers.  \cite{ofdm1,ofdm2,chirp} have  analyzed  various modulation techniques. There has also been work on underwater  channel estimation for OFDM~\cite{channel-ofdm1,channel-ofdm2}, the use of pilot symbols to track channel estimates within a packet~\cite{pilot-ofdm1,pilot-ofdm2},  Doppler estimation~\cite{doppler-ofdm1,doppler-ofdm2,doppler-ofdm3} and bit rate adaptation~\cite{rate1,rate3,rate4, rate5, rate6, rate7} for underwater sensor networks. Various hardware MAC protocols~\cite{mac-survey}  have also been explored in prior work.   There have also been interest in cognitive underwater spectrum access~\cite{underwater-cognitive1,underwater-cognitive2,underwater-cognitive3}. %While the PHY and MAC protocols presented in this paper builds on  these designs as well as in-air radio communication~\cite{kyle1,kyle2,sachin1,fara}
All this existing work has been designed for  custom underwater  hardware. In contrast, we  design  adaptation algorithms and    protocols that can run  on commodity mobile devices and  operate with a large diversity of  frequency responses  across  hardware.%  models. 

%including contention-free protocols like  FDMA, TDMA, CDMA~\cite{cdmamac,tdmamac} as well as contention-based protocols like Aloha, CSMA and RTS-CTS~\cite{aloha,csma,rtscts}. 

\noindent{\bf Underwater sensor networks/IoT.} 
Underwater sensor networks have been an active research area  and uses   custom sensor hardware~\cite{aquanet,book}.
 Recent work has designed novel  IoT hardware that uses backscatter communication~\cite{abc,wifibackscatter} and acoustic energy harvesting to design battery-free underwater wireless sensors~\cite{mit2,mit3}. Recent work has also designed  creative  hardware   to  communicate across the water and air interface~\cite{boundaries1,boundaries2}. {Optical approaches have also been proposed for short-range underwater communication at  1--2~m~\cite{liu2021uqcom}.} We build on this sensing work  but focus instead of using commodity devices  to enable underwater communication capabilities  without additional hardware. Further, in contrast to prior work that is designed for underwater sensors, our goal is to  enable humans to message underwater using their smart devices.

\noindent{\bf In-air acoustic communication.} Prior work has used smartphones to enable acoustic communication and tracking in air at close ranges~\cite{dhwani,acousticair1,acousticair2,acousticair3,millisonic,fingerio}. However enabling underwater communication using smartphones is challenging for multiple reasons, 1) multipath in underwater environments can be more severe in comparison to in air, 2) even the frequencies that can be used on the forward and backward paths can be  different underwater (see~\xref{sec:char}), 3) in-air acoustic   systems have a limited communication range of a few meters. Designing an underwater communication protocol that can adapt for a wide range of bit rates  and achieve much longer range  requires designing a different  system.

\begin{figure}[t!]
    \includegraphics[width=.3\textwidth]{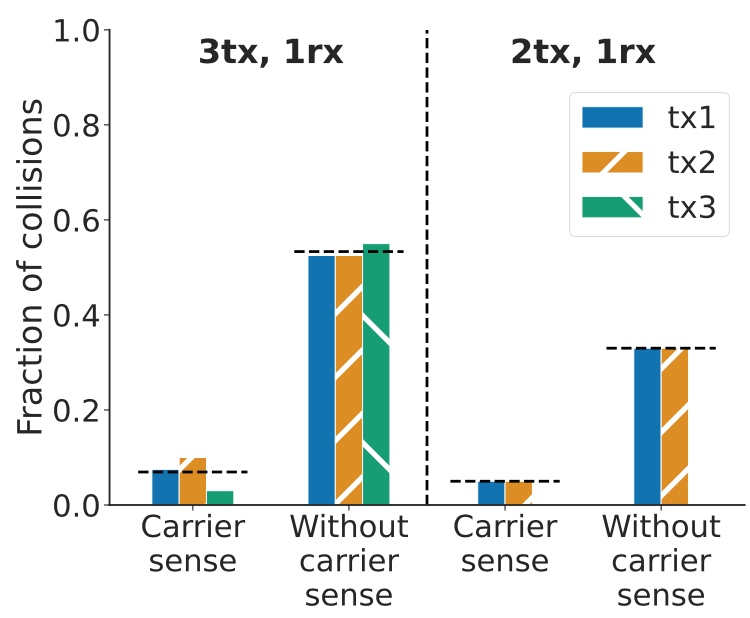}
    \vskip -0.15in
\caption{MAC protocol with multiple transmitters.}
\vskip -0.15in
\label{fig:mac}
\end{figure}

\section{Discussion}
We discuss various system-level aspects of our system.

\vskip 0.05in\noindent {{\bf Battery life.} To measure the power consumption of our system, we considered a setup with two  Samsung Galaxy S9 smartphones as the transmitter and the receiver. We ran our system continuously at maximum sound volume with the screen on and found that the battery power reduced by 32\% after a duration of 4 hours. This is sufficient for the application of recreational scuba diving, which have a maximum recommended dive time less than this~\cite{padi_table,padi2}.}

\vskip 0.05in\noindent{\bf Messaging latency.} A question that the reader may have is if the bit rates  in our design are sufficient for underwater messaging. Since typing underwater may be uncomfortable, our app interface has the user select from one of 240 messages, which translates to around 8 bits (and 12 bits after coding). It takes close to half a second to send this message  at 25 bps. At 1 kbps, we can even send   a 50 character message in  half a second. %, which is sufficient for our   application.

%\vskip 0.05in\noindent{\bf Message latency.} Our current system is designed for messaging and SoS applications that are important for both recreational and professional underwater activities. A future  direction to explore is to design a communication system that supports audio snippets or even low-resolution pictures. This may require using  compression techniques at the transmitter and  super-resolution algorithms at the receiver.

\vskip 0.05in\noindent{\bf  Range versus power.} The range achieved by custom hydrophones depends on their  transmit power. Commercial hydrophones that have a  kilometer or more range typically have a high transmit power~\cite{mit2}. The 30~m range for messaging and 100~m range for SoS beacons that we achieve with mobile devices is sufficient for many recreational and professional underwater activities. 

\vskip 0.05in\noindent{\bf Audibility.} Underwater acoustic modems  use a range of frequencies from multiple  kHz to 100s of kHz.  Our results show that 1-4~kHz is the optimal set of  frequencies for use on commodity smart phones and watches. These are in the audible range of human hearing, which is also true for some of the commercial  modems that operate in 7-17~kHz~\cite{lowfreq-modem,lowfreq-modem1}. Also note that human hearing underwater occurs through bone conduction rather than through the air pocket in the ear canal~\cite{bone1}. As a result, some prior studies observe that human hearing extends to ultrasonic frequencies underwater~\cite{bone1,bone2}. While fish and sharks have limited sensitivity for frequencies far above 10 kHz~\cite{fish1,fish2}, sea mammals, e.g.,
dolphins, seals, and whales  are highly sensitive to frequencies  up
to 150 kHz~\cite{whale1,whale2}. Transmissions from every acoustic modem are assumed to be audible to sea mammals in the vicinity~\cite{chirp}.

\section{Conclusion}
The last few decades have shown that software-based solutions can transform industries and bring technology to the masses more rapidly than custom hardware. We present the first acoustic system that uses software to bring underwater communication capabilities to commodity mobile devices. We believe that since our system  can be downloaded as a mobile software app on billions of devices, it can help democratize underwater communication and has the potential to be used by tens of  millions of people who participate in underwater activities like scuba divers and snorkeling every year.

\vskip 0.05in\noindent{\bf Acknowledgments.}
The researchers are funded by  the  Moore Inventor Fellow award \#10617 and the National Science Foundation. We thank our shepherd, Haitham Hassanieh, and the anonymous reviewers for their feedback on our submission.

\vskip 0.05in\noindent{This work does not raise any ethical issues.}

\bibliographystyle{ACM-Reference-Format}
\bibliography{underwatercomm}

\end{document}